\documentclass[preprint,11pt,authoryear]{elsarticle}
\usepackage[margin=1.25in]{geometry}
\usepackage{setspace}
\AtBeginDocument{
  \setlength{\abovedisplayskip}{8pt}
  \setlength{\belowdisplayskip}{8pt}
  \setlength{\abovedisplayshortskip}{6pt}
  \setlength{\belowdisplayshortskip}{6pt}
}

\usepackage{enumitem}
\setlist[enumerate,1]{label=(\roman*)} 
\usepackage{hyperref}
\usepackage{graphicx} 
\usepackage{xcolor}
\usepackage{amsmath}
\usepackage{amsthm}
\usepackage{mathtools}

\allowdisplaybreaks[4]
\usepackage{bm,amssymb}
\usepackage{bbm}
\usepackage{mathrsfs}
\usepackage{dsfont}
\usepackage{makecell,booktabs}
\usepackage{float}
\usepackage{multirow}
\usepackage{diagbox}
\usepackage{tabularx}
 \usepackage{threeparttable}
 \usepackage{xr}
\usepackage{subcaption}
\usepackage{algorithm,algpseudocode}
\usepackage[normalem]{ulem}
\newtheorem{theorem}{Theorem}
\newtheorem{assumption}{Assumption}
\newtheorem{proposition}{Proposition}

\newtheorem{example}{Example}

\usepackage{tikz}
\usetikzlibrary{positioning,arrows.meta,calc}

\definecolor{mygreen}{RGB}{0,140,100}

 \journal{xxx}
\begin{document}
\begin{frontmatter}
\title{Generalized optimal parameter-transfer learning through Mallows-type model averaging} 

\author[a]{Fen Jiang}
\author[b]{\texorpdfstring{Wenhui Li\corref{cor1}}{Wenhui Li}}
\ead{lwhui@amss.ac.cn}
\author[b,a]{Xinyu Zhang}

\address[a]{School of Management, University of Science and Technology of China, Hefei 230026, China}
\address[b]{State Key Laboratory of Mathematical Sciences, Academy of Mathematics and Systems Science, Chinese Academy of Sciences, Beijing 100190, China.}

\cortext[cor1]{Corresponding author}

\begin{abstract}
In many economic applications, multiple source datasets are available, but their effective combination is challenging due to heterogeneity across datasets. To address this problem, we study a parameter-transfer framework that shares only source-side estimates and propose a Mallows-type model averaging method for combining target and source models in the parametric setting. The weights are obtained from a Mallows-type criterion that is unbiased for the target prediction risk up to a weight-independent term, extending the classical Mallows criterion to the parameter-transfer framework. We establish that the proposed weights are asymptotically optimal when the target model is misspecified, and asymptotically allocate weights only to informative sources when the target model is correctly specified. These guarantees do not require any source model to be correctly specified. We also consider extensions of the framework to semiparametric and panel data settings. Simulation studies and house price application further demonstrate the effectiveness of our approach. 
\end{abstract}


\end{frontmatter}

\noindent\textbf{Running head:} Generalized Mallows Averaging for Transfer\\
\textbf{Proofs to:} Wenhui Li (\texttt{lwhui@amss.ac.cn})

\section{Introduction}

In many economic applications, data from a single domain may be inadequate for accurate prediction, while incorporating information from related datasets may improve predictive performance. For example, when predicting house prices in a given city with limited data, information from other cities with similar market conditions may be useful. However, heterogeneity across datasets makes direct pooling problematic. Transfer learning offers a principled framework for leveraging information from related source data to improve prediction on the target data \citep{pan2009survey}. This framework has been widely studied in statistics and econometrics, and a comprehensive review can be found in \citet{cao2023risk}.

Parameter-transfer is a class of transfer learning methods that improves estimation or prediction in a target domain by borrowing parameter information from related source domains. 
This is typically done under some form of similarity or shared structure between the source and target parameters. Parameter-transfer has been studied under a variety of model settings, including linear regression \citep{li2022transfer}, spatial econometric models \citep{zeng2026transfer}, and generalized linear models \citep{tian2023transfer}. Most of these methods follow a two-stage transfer paradigm: the first stage constructs an initial estimator using both source and target data, and the second stage refines or debiases it using the target data. Despite their success, these approaches have two important limitations. First, their theoretical guarantees are typically derived under correct model specification, whereas the true data-generating mechanism is unknown and model misspecification is common in practice. Second, many methods rely on pooled estimation or direct data sharing across domains, which may be infeasible due to privacy concerns or data access restrictions. In this paper, we develop a parameter-transfer approach based on the frequentist model averaging to address the above two issues.


A central question in transfer learning is how to effectively combine information from the target data and related source domains. In model-averaging-based transfer learning frameworks, this problem reduces to weight selection. Existing methods include Mallows-type approaches \citep{hansen2007least,hansen2009averaging,zhang2019inference}, cross-validation–based methods \citep{JMA2012,zhang2023model,li2025averaging,wei2025model,li2026factor}, and information-criterion–based methods \citep{buckland1997model,zhang2015model}. Among these, Mallows model averaging has become one of the most widely used approaches in the frequentist model-averaging literature and has been extended to a variety of complex models and data types \citep{zhu2019mallows,Zhang2020JASA,liao2021model}, making it a natural foundation for developing transfer learning methods. However, the theoretical development of Mallows-type methods in transfer learning settings remains limited, motivating our study.


Specifically, we propose a generalized Mallows-type parameter-transfer method, termed Generalized Transfer Learning based on Mallows Model Averaging (GTLMMA). Our approach aggregates source models through a weighted combination of their target predictions based on source-side parameter estimates, with weights determined by a Mallows-type criterion.



Our contributions are threefold. First, on the methodological side, we develop a generalized Mallows-type model averaging approach for transfer learning that extends the classical Mallows model averaging beyond linear models and nested model structures to a multi-source transfer learning framework. It applies to a broad class of parametric models, including linear and nonlinear regression, and spatial autoregressive models. In addition, unlike standard transfer-learning methods based on joint or pooled estimation using source and target data, the proposed approach uses source information only through source-side parameter estimates and does not require raw source data for target-side aggregation.

Second, on the theoretical side, we establish two key large-sample properties of GTLMMA, namely asymptotic optimality and weight consistency, both with explicit convergence rates. Under target-model misspecification, GTLMMA asymptotically achieves the minimal prediction risk and performs no worse than using the target data alone. Under correct specification, the weight consistency result shows that asymptotically nonzero weights are assigned only to an informative subset of models. Here, the informative set includes the target model and any source models that provide useful information for transfer, and may include misspecified yet predictive source models. In contrast, existing two-stage parameter-transfer methods (e.g., \citet{li2022transfer,tian2023transfer,zeng2026transfer}) are typically analyzed under correctly specified source models, while cross-validation-based transfer learning approaches (e.g., \citet{hu2023optimal,zhang2024prediction}) typically rely on stronger assumptions on how source and target parameters are related. Details on the construction and properties of the informative set are provided in Section~\ref{theo:weight}. Moreover, compared with cross-validation-based transfer learning methods (e.g., \citet{hu2023optimal, zhang2024prediction}), the proposed weight choice criterion is conditionally unbiased for the target prediction risk up to a weight-independent term, providing a theoretical basis for weight selection.

Third, on the computational and empirical side, the proposed method is computationally attractive compared with cross-validation-based transfer model averaging methods (e.g., \citet{hu2023optimal, zhang2024prediction}), as it avoids repeated refitting and does not require selecting tuning parameters such as the number of folds. We further demonstrate its effectiveness through extensive simulation studies and empirical applications.

The structure of the paper is as follows.
Section \ref{framework} introduces the model framework and the proposed GTLMMA method, with illustrative parametric specifications. 
Section \ref{theory} establishes the large sample properties, including asymptotic optimality and weight consistency. 
Sections \ref{sim} and \ref{Empirical} report simulation results and an empirical application to house-price prediction. 
Section \ref{exten} presents extensions to semiparametric, panel data, and nonparametric settings.
Section \ref{Discussion} offers concluding remarks and potential directions for future research. The appendices contain additional calculations, discussions of assumptions, and details on model specification. The supplementary material provides a supporting lemma, proofs of the main results, and a semiparametric application.

\vspace{0.3cm}
\noindent\textbf{Notations.} \quad Denote the set of all candidate model indices by $\mathcal{K}=\{0,\dots,K\}$, and the minimum sample size by $\underline{n}=\min_{k \in \mathcal{K}}{n_k}$. We use $A \setminus B$ to denote the set difference for sets $A$ and $B$. Define $\bm{I}_{n}$ as an $n \times n$ identity matrix. Let $\lambda_{\min}(\cdot)$ and $\lambda_{\max}(\cdot)$ be the minimum and maximum eigenvalues of a given matrix. In addition, let $\sigma_{\max}(\cdot)$ denote the largest singular value of a matrix. Let $\|\cdot\|$ denote the $\ell_2$-norm when applied to vectors, and the spectral norm when applied to matrices. Let $a \vee b$ denote the maximum of $a$ and $b$. The notation $a_n \gg b_n$ means that $a_n$ increases at a rate faster than $b_n$ as $\underline{n} \to \infty$. Let $a_n \asymp b_n$ denote that $|a_n / b_n| \to c$ for some positive constant $c$ as $\underline{n} \to \infty$.

\section{Model setting and parameter-transfer methodology} \label{framework}
\subsection{Model setting}\label{modelsetting}
Let $\{(\bm{x}_{i}^{(0)},y_{i}^{(0)})\}_{i=1}^{n_0}$ denote the target sample drawn from the target population, where $\bm{x}_{i}^{(0)}$, $i=1,\dots,n_0$ are $p_0$-dimensional independent and identically distributed (i.i.d.) observations. Denote the design matrix of exogenous variables as $\bm{X}^{(0)} =(\bm{x}_{1}^{(0)},\dots,\bm{x}_{n_0}^{(0)})^{\top}$, and the vector of dependent variables as $\bm{y}^{(0)}=( y_{1}^{(0)},\dots,y_{n_0}^{(0)})^{\top}$. Given $\bm{X}^{(0)}$, $\bm{y}^{(0)}$ is generated from the following model:
\begin{align*}
    \bm{y}^{(0)} = \bm{\mu}^{(0)} + \bm{e}^{(0)},
\end{align*} 
where $\bm{\mu}^{(0)}=\mathbb{E}(\bm{y}^{(0)} \mid \bm{X}^{(0)})$ and $\bm{e}^{(0)} = (e_1^{(0)}, \dots, e_{n_0}^{(0)})^{\top}$ represents the vector of stochastic disturbances. Define $\bm{\Omega}_0=\mathrm{Var}(\bm{y}^{(0)} \mid \bm{X}^{(0)})$ as the conditional covariance matrix of $\bm{y}^{(0)}$, and let $\widehat{\bm{\Omega}}_0$ denote its estimator.

Our objective is to predict the conditional mean $\bm{\mu}^{(0)}$. When the sample size $n_0$ is small, the predictive performance may be unsatisfactory. To improve prediction, we adopt transfer learning techniques to leverage information from related source datasets. Suppose that we have source samples from $K$ potentially heterogeneous populations. For $k=1,\dots,K$, let $\{(\bm{x}_{i}^{(k)},y_{i}^{(k)})\}_{i=1}^{n_k}$ denote the $k$th group of source samples from the $k$th population, where $\bm{x}_{i}^{(k)}$, $i=1,\dots,n_k$ are $p_k$-dimensional i.i.d. observations. For $k=1,\dots,K$, denote the design matrix of exogenous variables as $\bm{X}^{(k)} =(\bm{x}_{1}^{(k)},\dots,\bm{x}_{n_k}^{(k)})^{\top}$, and the vector of dependent variables as $\bm{y}^{(k)}=( y_{1}^{(k)},\dots,y_{n_k}^{(k)})^{\top}$. For $k=1,\dots,K$, given $\bm{X}^{(k)}$, $\bm{y}^{(k)}$ is generated from the following model:
\begin{align*}
    \bm{y}^{(k)} = \bm{\mu}^{(k)} + \bm{e}^{(k)},
\end{align*} 
where $\bm{\mu}^{(k)}=\mathbb{E}(\bm{y}^{(k)} \mid \bm{X}^{(k)})$, and $\bm{e}^{(k)} = (e_1^{(k)}, \dots, e_{n_k}^{(k)})^{\top}$ represents the vector of stochastic disturbances. In the transfer learning framework, we assume that $\bm{e}^{(0)}, \dots, \bm{e}^{(K)}$ are conditionally independent given $\mathcal{X}$ with $\mathcal{X}=\{\bm{X}^{(0)},\dots,\bm{X}^{(K)}\}$. We also assume that, given $\bm{X}^{(k)}$, the conditional distribution of $\bm{e}^{(k)}$ is independent of $\mathcal{X} \setminus \{\bm{X}^{(k)}\}$ for any $k=0,\dots,K$.  

We construct a transfer learning framework that leverages information from $\bm{\mu}^{(k)}$, $k=1,\dots,K$, to improve the prediction of $\bm{\mu}^{(0)}$. Since the true conditional mean functions $\bm{\mu}^{(k)}$ are generally unknown, we approximate them by working models, denoted by $\bm{\mu}^{(k)}_{\mathrm{wm}}$. In the parametric setting, $\bm{\mu}^{(k)}_{\mathrm{wm}}$ is specified through a  parameter $\bm{\delta}^{(k)}$, and several representative specifications are presented in Section \ref{example}.

\subsection{GTLMMA}\label{method}
We now introduce the GTLMMA method, which combines predictions from the target and source models through data-driven weighting. Let $\widehat{\bm{\mu}}^{(k)}$, $k=0,\dots,K$, denote candidate predictions of $\bm{\mu}^{(0)}$. Define the model averaging estimator as $\widehat{\bm{\mu}}(\bm{\omega}) = \sum_{k=0}^K \omega_k\widehat{\bm{\mu}}^{(k)}$, where $\bm{\omega}=(\omega_0,\dots,\omega_K)^{\top}$ is a weight vector in the weight set $\mathcal{W}=\{ \bm{\omega} \in [0,1]^{K+1} : {\textstyle \sum_{k=0}^{K}}\omega_k=1 \}$. Assuming that $\bm{\Omega}_0$ is known, we propose the following Mallows-type criterion to choose an appropriate weight vector $\bm{\omega}$ for $\widehat{\bm{\mu}}(\bm{\omega})$:
\begin{align*}
\mathcal{C}(\bm{\omega})=\|  \bm{y}^{(0)} - \widehat{\bm{\mu}}(\bm{\omega}) \|^2 + 2 \omega_0 \mathrm{tr}   \left\{ \partial \widehat{\bm{\mu}}^{(0)} / \partial (\bm{y}^{(0)})^{\top} \bm{\Omega}_0 \right\},
\end{align*}
where $\partial$ denotes the partial differential operator. The first term of $\mathcal{C}(\bm{\omega})$  measures the goodness-of-fit of the model averaging estimator. The second term of $\mathcal{C}(\bm{\omega})$ can be shown to be a conditionally unbiased estimator of $2df(\bm{\omega})$, where $df(\bm{\omega})$ denotes the prediction degrees of freedom of the target model, defined as $df(\bm{\omega})=\mathrm{tr}[\mathrm{cov}\{\widehat{\bm{\mu}}(\bm{\omega}),\bm{y}^{(0)} \mid \mathcal{X} \}]$, following \citet{Hodges2001}, \citet{LH2008}, and \citet{Z&Y2018}. Therefore, $\mathcal{C}(\bm{\omega})$ can be regarded as a generalization of the classical Mallows criterion \citep{hansen2007least}. The computation of 
$\partial \widehat{\bm{\mu}}^{(0)} / \partial (\bm{y}^{(0)})^{\top}$ 
across different classes of predictors is discussed in \ref{cau:derivative}.

Define the squared prediction loss and the corresponding prediction risk as $L(\bm{\omega}) =\| \widehat{\bm{\mu}}(\bm{\omega}) - \bm{\mu}^{(0)} \|^2$ and $R(\bm{\omega}) = \mathbb{E}\{ L(\bm{\omega}) \mid \mathcal{X}\} = \mathbb{E}( \| \widehat{\bm{\mu}}(\bm{\omega}) - \bm{\mu}^{(0)} \|^2 \mid \mathcal{X})$, respectively. We now turn to a theorem showing that $\mathcal{C}(\bm{\omega})$ is conditionally unbiased for the risk $R(\bm{\omega})$ up to a term. Before stating the theorem, we first introduce some standard conditions. 

\begin{assumption}\label{assumption1}
\leavevmode
    \begin{enumerate}
        \item The error $\bm{e}^{(0)}$ is normally distributed given $\bm{X}^{(0)}$. In addition, $\widehat{\bm{\mu}}^{(0)}$ is almost differentiable in $\bm{y}^{(0)}$, with the regularity conditions for Stein’s lemma \citep{stein1981} satisfied. 
        \item $\widehat{\bm{\mu}}^{(0)} = \bm{Q}_0 \bm{y}^{(0)}$, where $\bm{Q}_0$ depends only on $\bm{X}^{(0)}$.
    \end{enumerate}
\end{assumption}

Assumptions \ref{assumption1}(i) and \ref{assumption1}(ii) are standard and typically mild. The normality condition in Assumption \ref{assumption1}(i) is commonly adopted in the model averaging literature (e.g., \citet{zhang2014biometrika}, \citet{zhang2015model}, \citet{Z&Y2018}). The differentiability condition in Assumption~\ref{assumption1}(i) ensures the applicability of Stein’s lemma, which is used to characterize the prediction degrees of freedom. Assumption \ref{assumption1}(ii) requires that the predictor is linear in $\bm{y}^{(0)}$, with $\bm{Q}_0$ depending only on $\bm{X}^{(0)}$. This condition is characteristic of linear smoothers (see \citet{hastie2009elements}) and is satisfied by many machine learning and econometric methods, as illustrated in Eq. (11) of \citet{qiu2023boosting}. Some illustrative examples satisfying Assumption~\ref{assumption1}(ii) are provided in Section~\ref{example}. Then, we demonstrate the unbiasedness of $\mathcal{C}(\bm{\omega})$ in the following theorem.

\begin{theorem}\label{theorem1}
Under Assumption \ref{assumption1}(i), or alternatively under Assumption \ref{assumption1}(ii), the criterion $\mathcal{C}(\bm{\omega})$ is a conditionally unbiased estimator of the risk $R(\bm{\omega})$ up to an additive term, i.e., $\mathbb{E}\{\mathcal{C}(\bm{\omega}) \mid \mathcal{X}\} =R(\bm{\omega}) +  \mathrm{tr}(\bm{\Omega}_0)$.
\end{theorem}

Theorem \ref{theorem1} implies that, conditional on $\mathcal{X}$, minimizing the proposed Mallows-type criterion is equivalent to minimizing the target risk $R(\bm{\omega})$. 
As a consequence, the criterion is directly aligned with the objective of improving performance on the target predictive task, and the resulting weights provide a coherent mechanism for incorporating source information into target prediction.

If $\bm{\Omega}_0$ is unknown, we replace it with its estimator $\widehat{\bm{\Omega}}_0$ in $\mathcal{C}(\bm{\omega})$ and derive
\begin{align}\label{eq4}
\widehat{\mathcal{C}}(\bm{\omega})=\|  \bm{y}^{(0)} - \widehat{\bm{\mu}}(\bm{\omega}) \|^2 + 2 \omega_0 \mathrm{tr}   \left\{  \partial \widehat{\bm{\mu}}^{(0)} / \partial (\bm{y}^{(0)})^{\top} \widehat{\bm{\Omega}}_0 \right\}.
\end{align}
By minimizing $\widehat{\mathcal{C}}(\bm{\omega})$, we obtain the optimal weight $\widehat{\bm{\omega}}=\underset{\bm{\omega} \in \mathcal{W} }{\arg\min} \,\widehat{\mathcal{C}}(\bm{\omega})$. Substituting $\bm{\omega}$ by $\widehat{\bm{\omega}}$ in $\widehat{\bm{\mu}}(\bm{\omega})$ results in the GTLMMA prediction $\widehat{\bm{\mu}}(\widehat{\bm{\omega}})$. The data flow of the GTLMMA method is illustrated in Figure~\ref{fig:GTLMMA_flow}.

Next, we show that \eqref{eq4} is a quadratic function of $\bm{\omega}$. Let $\bm{D}$ be an $n_0 \times (K+1)$ matrix whose $j$th column is $\bm{y}^{(0)}- \widehat{\bm{\mu}}^{(j-1)}$ for $j=1,\dots,K+1$. Let $\bm{d}$ be a $(K+1) \times 1$ vector whose first element is $ 2 \mathrm{tr} \{ \partial \widehat{\bm{\mu}}^{(0)}/\partial (\bm{y}^{(0)})^{\top} \widehat{\bm{\Omega}}_0 \} $ and all other elements are zeros. Then, it follows that 
\begin{eqnarray*}
\widehat{\mathcal{C}}(\bm{\omega})=
\bm{\omega}^{\top}\bm{D}^{\top}\bm{D}\bm{\omega}+\bm{d}^{\top}\bm{\omega}.
\end{eqnarray*}
Therefore, minimizing $\widehat{\mathcal{C}}(\bm{\omega})$ over the weight set $\mathcal{W}$ is a constrained quadratic programming problem, which can be efficiently solved by common R packages.

\begin{figure}[t]
\centering
\begin{tikzpicture}[
    scale=0.85,
    transform shape,
    >=Latex,
    font=\footnotesize,
    dataTarget/.style={draw=black, rounded corners, align=center, minimum width=2.15cm, minimum height=0.62cm},
    dataSource/.style={draw=black, rounded corners, align=center, minimum width=2.15cm, minimum height=0.62cm},
    box/.style={draw=black, rounded corners, align=center, minimum width=1.75cm, minimum height=0.55cm},
    predBox/.style={draw=black, rounded corners, align=center, minimum width=2.05cm, minimum height=0.62cm},
    mergeBox/.style={draw=black, rounded corners, align=center, minimum width=4.25cm, minimum height=0.62cm},
    weightBox/.style={draw=black, rounded corners, align=center, minimum width=5.10cm, minimum height=0.78cm},
    finalBox/.style={draw=black, rounded corners, align=center, minimum width=4.35cm, minimum height=0.68cm},
    arrow/.style={->, thin, black},
    dashedArrow/.style={->, thin, dashed, black},
    guide/.style={dashed, gray!45}
]

\def\xshift{0.55}

\node[align=left, font=\bfseries] at (-5.1, 4.0) {Input data};
\node[align=left, font=\bfseries] at (-5.1, 1.45) {Step 1:\\Construct\\predictions};
\node[align=left, font=\bfseries] at (-5.1, -2.70) {Step 2:\\Weight\\selection};
\node[align=left, font=\bfseries] at (-5.1, -4.50) {Step 3:\\Weighted\\prediction};

\node[dataTarget] (targetData) at (-3.0+\xshift, 4.0)
{Target data\\$(\bm X^{(0)},\bm y^{(0)})$};

\node[dataSource] (source1Data) at (-0.2+\xshift, 4.0)
{Source data 1\\$(\bm X^{(1)},\bm y^{(1)})$};

\node[dataSource] (source2Data) at (2.6+\xshift, 4.0)
{Source data 2\\$(\bm X^{(2)},\bm y^{(2)})$};

\node[font=\normalsize] (dotsData) at (4.7+\xshift, 4.0) {$\cdots$};

\node[dataSource] (sourceKData) at (6.9+\xshift, 4.0)
{Source data $K$\\$(\bm X^{(K)},\bm y^{(K)})$};

\node[box] (targetFit) at (-3.0+\xshift, 2.35) {Fit target\\model};
\node[box] (source1Fit) at (-0.2+\xshift, 2.35) {Fit source\\model 1};
\node[box] (source2Fit) at (2.6+\xshift, 2.35) {Fit source\\model 2};
\node[box] (sourceKFit) at (6.9+\xshift, 2.35) {Fit source\\model $K$};

\node[predBox, text width=2.0cm] (targetPred) at (-3.0+\xshift, 0.75)
{Target prediction\\$\widehat{\bm\mu}^{(0)}$};

\node[predBox, text width=2.0cm] (source1Pred) at (-0.2+\xshift, 0.75)
{Prediction 1 on $\bm X^{(0)}$\\$\widehat{\bm\mu}^{(1)}$};

\node[predBox, text width=2.0cm] (source2Pred) at (2.6+\xshift, 0.75)
{Prediction 2 on $\bm X^{(0)}$\\$\widehat{\bm\mu}^{(2)}$};

\node[font=\normalsize] (dotsPred) at (4.7+\xshift, 0.75) {$\cdots$};

\node[predBox, text width=2.0cm] (sourceKPred) at (6.9+\xshift, 0.75)
{Prediction $K$ on $\bm X^{(0)}$\\$\widehat{\bm\mu}^{(K)}$};

\node[mergeBox] (candidate) at (1.9+\xshift, -1.25)
{Candidate predictions $\{\widehat{\bm\mu}^{(k)}\}_{k=0}^{K}$};

\node[weightBox] (weight) at (1.9+\xshift, -2.70)
{Weight selection via the GTLMMA criterion $\widehat{\mathcal{C}}(\bm\omega)$};

\node[finalBox] (final) at (1.9+\xshift, -4.50)
{GTLMMA prediction $\widehat{\bm\mu}(\widehat{\bm\omega})
=\sum_{k=0}^{K}\widehat{\omega}_k\widehat{\bm\mu}^{(k)}$};

\draw[arrow] (targetData) -- (targetFit);
\draw[arrow] (source1Data) -- (source1Fit);
\draw[arrow] (source2Data) -- (source2Fit);
\draw[arrow] (sourceKData) -- (sourceKFit);

\draw[arrow] (targetFit) -- (targetPred);
\draw[arrow] (source1Fit) -- (source1Pred);
\draw[arrow] (source2Fit) -- (source2Pred);
\draw[arrow] (sourceKFit) -- (sourceKPred);

\draw[arrow] (targetPred.south) -- ++(0,-0.55) -| ([xshift=-1.55cm]candidate.north);
\draw[arrow] (source1Pred.south) -- ++(0,-0.55) -| ([xshift=-0.50cm]candidate.north);
\draw[arrow] (source2Pred.south) -- ++(0,-0.55) -| ([xshift=0.50cm]candidate.north);
\draw[arrow] (sourceKPred.south) -- ++(0,-0.55) -| ([xshift=1.55cm]candidate.north);

\draw[arrow] (candidate) -- (weight);

\draw[dashedArrow] (targetData.west) -- ++(-0.25,0) |- (weight.west);

\draw[arrow] (weight) -- (final);

\draw[guide] (-5.4, 3.15) -- (8.6, 3.15);
\draw[guide] (-5.4, -1.85) -- (8.6, -1.85);
\draw[guide] (-5.4, -3.55) -- (8.6, -3.55);

\end{tikzpicture}
\caption{Data flow of the GTLMMA method with weight selection via the GTLMMA criterion.}
\label{fig:GTLMMA_flow}
\end{figure}


\subsection{Illustrative parametric specifications}\label{example}

In this section, we present several commonly used parametric model specifications and derive the corresponding expressions for $\bm{\mu}^{(k)}_{\mathrm{wm}}$ and $\partial \widehat{\bm{\mu}}^{(0)}/\partial (\bm{y}^{(0)})^{\top}$ required for constructing the GTLMMA criterion in \eqref{eq4}.

\setcounter{example}{0}

We assume $p_0 = \dots = p_K$, denoting this common dimension by $p$. Under the parametric setting, the mean function $\bm{\mu}^{(k)}$ is modeled through a parametric form $f(\bm{X}^{(k)}, \bm{\delta}^{(k)})$. Therefore, transferring information across domains can be viewed as transferring the parameters $\bm{\delta}^{(k)}$ in the parametric setting.

   \begin{example}{ \textbf{Linear regression model}}

   A commonly used working model is the linear regression specification $\bm{\mu}^{(k)}_{\mathrm{wm}} = \bm{X}^{(k)} \bm{\delta}^{(k)}$, where $\bm{\delta}^{(k)} \in \mathbb{R}^{p}$. For the target sample, the prediction is $\widehat{\bm{\mu}}^{(0)}= \bm{X}^{(0)} \widehat{\bm{\delta}}^{(0)}$, with derivative $\partial \widehat{\bm{\mu}}^{(0)}/\partial (\bm{y}^{(0)})^{\top}= \bm{X}^{(0)} \{ \partial \widehat{\bm{\delta}}^{(0)}/\partial(\bm{y}^{(0)})^{\top} \}$. Under ordinary least squares, this derivative coincides with the hat matrix $\bm{X}^{(0)} \{ (\bm{X}^{(0)})^{\top} \bm{X}^{(0)}\}^{-1} (\bm{X}^{(0)})^{\top}$, while under ridge regression with penalty $\lambda$, it becomes $\bm{X}^{(0)} \{(\bm{X}^{(0)})^{\top} \bm{X}^{(0)} + \lambda \bm{I}_p\}^{-1} (\bm{X}^{(0)})^{\top}$.
   \end{example}

   \begin{example}{ \textbf{Nonlinear regression model}}

   We consider a nonlinear regression specified as $\bm{\mu}^{(k)}_{\mathrm{wm}} = [\phi(\bm{x}_1^{(k)},\bm{\delta}^{(k)}),\dots,\phi(\bm{x}_{n_k}^{(k)},\bm{\delta}^{(k)})]^{\top}$, where $\phi(\cdot)$ is a known nonlinear function and $\bm{\delta}^{(k)} \in \Theta_k \subset \mathbb{R}^{p}$. The parameter $\bm{\delta}^{(k)}$ can be estimated, for example, by nonlinear least squares under standard regularity conditions \citep{Jennrich1969}. For simplicity, we assume homoskedastic target errors with $\mathbb{E}(e_i^{(0)} \mid \bm{x}_i^{(0)}) = 0$ and $\mathrm{Var}(e_i^{(0)} \mid \bm{x}_i^{(0)}) = \sigma_0^2$. Under this setting, we obtain $\mathrm{tr} \{  \partial \widehat{\bm{\mu}}^{(0)}/\partial (\bm{y}^{(0)})^{\top} \widehat{\bm{\Omega}}_0 \}=\widehat{\sigma}_0^2 \sum_{i=1}^{n_0} \partial \widehat{\mu}_i^{(0)}/\partial y_i^{(0)}$, where $\widehat{\sigma}_0^2$ denotes an estimator of $\sigma_0^2$. The detailed derivation is provided in \ref{calcu:nonli}.
   
    \end{example}

    \begin{example}{ \textbf{Spatial autoregressive (SAR) model}} 

    Spatial dependence is commonly modeled using the spatial autoregressive (SAR) specification \citep{cli1973spatial, anselin2010thirty}. For $k=0,\dots,K$, consider the SAR model $\bm{y}^{(k)} = \rho^{(k)} \bm{W}_k \bm{y}^{(k)} + \bm{X}^{(k)}\bm{\beta}^{(k)} + \bm{\epsilon}^{(k)}$, where $\bm{W}_k$ is a known non-stochastic spatial weights matrix with zero diagonals, $\rho^{(k)}$ is the spatial dependence parameter, and the random error $\bm{\epsilon}^{(k)}$ satisfying $\mathbb{E}( \bm{\epsilon}^{(k)} \mid \bm{X}^{(k)})= \bm{0}$ and $\mathrm{Var}\{\bm{\epsilon}^{(k)} \mid \bm{X}^{(k)}\}=\sigma_k^2\bm{I}_{n_k}$. If the matrix $(\bm{I}_{n_k} - \rho^{(k)} \bm{W}_k)$ is invertible, the implied working mean is  $\bm{\mu}^{(k)}_{\mathrm{wm}} = (\bm{I}_{n_k} - \rho^{(k)} \bm{W}_k)^{-1} \bm{X}^{(k)}\bm{\beta}^{(k)}$, with parameter vector $\bm{\delta}^{(k)}=\{ (\bm{\beta}^{(k)})^{\top},\rho^{(k)} \}^{\top} \in \mathbb{R}^{p+1}$. For each $k=0,\dots,K$, the fitted model is evaluated on the target design matrix $\bm{X}^{(0)}$ to produce $\widehat{\bm{\mu}}^{(k)} \in \mathbb{R}^{n_0}$. For the SAR model, this yields $\widehat{\bm{\mu}}^{(k)}=(\bm{I}_{n_0} -  \widehat{\rho}^{(k)} \bm{W}_0)^{-1} \bm{X}^{(0)} \widehat{\bm{\beta}}^{(k)}$, where $\widehat{\rho}^{(k)}$ and $\widehat{\bm{\beta}}^{(k)}$ are estimators of the spatial and regression parameters.
    Let $\widehat{\bm{S}}_0=\bm{I}_{n_0} - \widehat{\rho}^{(0)} \bm{W}_0$. Then $\partial \widehat{\bm{\mu}}^{(0)}/\partial (\bm{y}^{(0)})^{\top}=\widehat{\bm{S}}_0^{-1} \left\{  \bm{W}_0  \widehat{\bm{S}}_0^{-1} \bm{X}^{(0)}  \widehat{\bm{\beta}}^{(0)} \partial \widehat{\rho}^{(0)}/\partial (\bm{y}^{(0)})^{\top} +  \bm{X}^{(0)} \partial \widehat{\bm{\beta}}^{(0)}/\partial (\bm{y}^{(0)})^{\top} \right\}$. For simplicity, we assume that the spatial weights matrices for the target and source models share a common construction scheme. The explicit form of $\partial \widehat{\bm{\mu}}^{(0)}/\partial (\bm{y}^{(0)})^{\top}$ for the SAR model is provided in \ref{calcu:sar}. Section \ref{add:dis:sarar} of the supplementary material provides an intuitive discussion of parameter-transfer under the SAR model, along with alternative constructions of spatial weight matrices and extensions to more general spatial models.
    \end{example}

\section{Large sample properties}\label{theory}

This section studies the large sample properties of GTLMMA. For a misspecified target model, we discuss the asymptotic optimality of GTLMMA. For a correctly specified target model, we focus on the weight consistency. A model is said to be correctly specified if the working model $\bm{\mu}^{(k)}_{\mathrm{wm}}$ coincides with the true data-generating function $\bm{\mu}^{(k)}$, and misspecified otherwise. Model misspecification is common and may arise from various factors, such as incorrect functional forms or omitted variables.

All limiting processes discussed in this section are with respect to $\underline{n} \to \infty$. The number of candidate models $K$ and the parameter dimension $p$ are allowed to diverge with $\underline{n}$ under $p < \underline{n}$.

\subsection{Asymptotic optimality}\label{theo:opti}

In this section, we discuss the in-sample and out-of-sample asymptotic optimality of GTLMMA. First, we discuss the asymptotic optimality of GTLMMA for in-sample prediction. The required assumptions are as follows.

\begin{assumption}\label{assumption2}
\leavevmode
\begin{enumerate}
    \item $\mathbb{E}\{n_0^{-1}\|\bm{\mu}^{(0)}\|^2\}=O(1)$ and $\mathbb{E}\{n_0^{-1}\|\bm{e}^{(0)}\|^2\}= O(1)$.
    \item $\lambda_{\max}(\bm{\Omega}_0)\le c_0$ almost surely for some constant $c_0>0$, and the covariance estimator $\widehat{\bm{\Omega}}_0$ satisfies $\sigma_{\max}(\widehat{\bm{\Omega}}_0)=O_p(1)$.
\end{enumerate}
\end{assumption}

Assumption~\ref{assumption2}(i) requires that the second moments of $\bm{\mu}^{(0)}$ and $\bm{e}^{(0)}$ are of order $O(1)$, which is similar to Assumption 2(i) of \citet{zhang2023model}, Eq.~(11) of \citet{Wan2010}, and Assumption 7 of \citet{liu2016generalized}. Moreover, it imposes weaker moment conditions on the error term $\bm{e}^{(0)}$ than those required in the existing Mallows-type literature (e.g., \citet{b24}, \citet{hansen2007least}, and \citet{Wan2010}). Assumption~\ref{assumption2}(ii) requires that the covariance matrix $\bm{\Omega}_0$ and its estimator $\widehat{\bm{\Omega}}_0$ are bounded in spectral norm, which is analogous to Assumption 2 of \citet{Z&Y2018}.

\begin{assumption}\label{assumption3}
$\mathrm{rank}\{ \partial \widehat{\bm{\mu}}^{(0)}/\partial (\bm{y}^{(0)})^{\top}\}=O_p(p^{1/2}n_0^{1/2})$ and $\sigma_{\max}\{ \partial \widehat{\bm{\mu}}^{(0)}/\partial (\bm{y}^{(0)})^{\top}\}=O_p(1)$.
\end{assumption}

Assumption~\ref{assumption3} imposes mild structural regularity conditions on the Jacobian matrix $\partial \widehat{\bm{\mu}}^{(0)} /\partial (\bm{y}^{(0)})^{\top}$. It allows the rank of the Jacobian is of order $O_p(p^{1/2}n_0^{1/2})$, while requiring its largest singular value to remain bounded. The implications of Assumption \ref{assumption3} for different models will be discussed in detail in \ref{dis:ass}.

\begin{assumption}\label{assumption4}
There exist limiting values $\bm{\delta}^{(k)*}$  such that $K^{-1/2}\alpha_k^{-1/2}\|\widehat{\bm{\delta}}^{(k)}-\bm{\delta}^{(k)*} \|=O_p(1)$ 
uniformly for all $k \in \mathcal{K}$, with $\alpha_k$ depending on the sample size $n_k$ and satisfying $\alpha_k \ge p n_k^{-1}$.
\end{assumption}

Assumption \ref{assumption4} concerns the convergence rate of $\widehat{\bm{\delta}}^{(k)}$ to its limiting value $\bm{\delta}^{(k)*}$, which is a common requirement in the literature (e.g., \citet{zhang2023model,hu2023optimal,zhang2024prediction}). When the $k$th model is correctly specified, the limiting value $\bm{\delta}^{(k)*}$ equals the true parameter value $\bm{\delta}^{(k)}$. In contrast, when the model is misspecified, the limiting value $\bm{\delta}^{(k)*}$ is a pseudo-true value, which minimizes the Kullback-Leibler Information Criterion \citep{white1982maximum}. Since the number of source models $K$ is allowed to diverge, we modify the convergence rate requirement compared to \citet{white1982maximum} to ensure uniform convergence across all models. Moreover, since the parameter dimension $p$ is also allowed to diverge with the sample size, the rate parameter $\alpha_k$ may depend on the sample size $n_k$, with a common choice being $p n_k^{-1}$ (e.g., Condition~3 of \citet{hu2023optimal}). In contrast, Assumption~\ref{assumption4} does not fix $\alpha_k$ to $p n_k^{-1}$ but instead formulates the rate condition in terms of $\alpha_k^{1/2}K^{1/2}$, thereby permitting slower convergence than $p^{1/2} n_k^{-1/2} K^{1/2}$. This relaxation allows the assumption to cover a broader class of estimation methods, including, for example, nonlinear least-squares estimators \citep{POLLARD2006548}. Although this formulation allows for slower convergence, we require $K^{1/2}\bar{\alpha}^{1/2} \to 0$ to ensure that the estimation error remains asymptotically negligible, where $\bar{\alpha}=\sup_{k \in \mathcal{K}}{\alpha_k}$.


\begin{assumption}\label{assumption5} 
For $k =0,\dots,K$, define the local neighborhood of $\bm{\delta}^{(k)*}$ as $\bm{\Delta}^{(k)}= \{\overline{\bm{\delta}}^{(k)}: \alpha_k^{-1/2} K^{-1/2} \|\overline{\bm{\delta}}^{(k)}-\bm{\delta}^{(k)*} \|< c_1\}$, where $c_1$ is a positive constant. For any $\overline{\bm{\delta}}^{(k)} \in \bm{\Delta}^{(k)}$, let $\overline{\bm{\mu}}^{(k)}=f(\bm{X}^{(0)},\overline{\bm{\delta}}^{(k)})$, and assume that $\overline{\bm{\mu}}^{(k)}$ is differentiable with respect to $\overline{\bm{\delta}}^{(k)}$ on $\bm{\Delta}^{(k)}$. Moreover, $\sup_{\overline{\bm{\delta}}^{(k)} \in \bm{\Delta}^{(k)}} \| \partial \overline{\bm{\mu}}^{(k)} / \partial (\bm{\overline{\delta}}^{(k)})^{\top} \|^2=O_p(n_0)$ uniformly for $k \in \mathcal{K}$.
\end{assumption}

Assumption \ref{assumption5} is a relatively mild constraint concerning the differentiability of $\overline{\bm{\mu}}^{(k)}$. Define $\bm{U}_k=\partial \overline{\bm{\mu}}^{(k)}/ \partial (\bm{\overline{\delta}}^{(k)})^{\top}$. Since $\|\bm{U}_k\|^2=\lambda_{\max}(\bm{U}_k^{\top}\bm{U}_k)$, \ref{dis:ass}  examines the structure of $\lambda_{\max}(\bm{U}_k^{\top}\bm{U}_k)$ and provides sufficient conditions ensuring Assumption \ref{assumption5} under different scenarios.

Next, we define some limiting quantities when the parameter estimates are substituted by their corresponding pseudo-true values, such as $\widehat{\bm{\mu}}^{(k)}$ by $\widetilde{\bm{\mu}}^{(k)}=f(\bm{X}^{(0)},\bm{\delta}^{(k)*})$, $\widehat{\bm{\mu}}(\bm{\omega})$ by $\widetilde{\bm{\mu}}(\bm{\omega})=\sum_{k=0}^K \omega_k \widetilde{\bm{\mu}}^{(k)}$, and the risk $R(\bm{\omega})$ by $\widetilde{R}(\bm{\omega})=\mathbb{E} ( \| \widetilde{\bm{\mu}}(\bm{\omega})-\bm{\mu}^{(0)}\|^2  \mid  \bm{X}^{(0)} )$. Define $\widetilde{\xi}=\inf_{ \bm{\omega} \in \mathcal{W} } \widetilde{R}(\bm{\omega})$. Relevant restrictions are shown in Assumption \ref{assumption6}.

\begin{assumption}\label{assumption6} 
\leavevmode
\begin{enumerate}
    \item For $k \in \mathcal{K}$, $\mathbb{E}\{n_0^{-1}\|\widetilde{\bm{\mu}}^{(k)}\|^2\} = O(1)$.
    \item $\widetilde{\xi}^{-1} \sup_{\bm{\omega} \in \mathcal{W}} | \| \widehat{\bm{\mu}}(\bm{\omega})-\bm{\mu}^{(0)}\|^2 -\|\widetilde{\bm{\mu}}(\bm{\omega})-\bm{\mu}^{(0)}\|^2 | $ is uniformly integrable. 
    \item $\widetilde{\xi}^{-1} \bar{\alpha}^{1/2} Kn_0=o(1)$ almost surely.
\end{enumerate}
\end{assumption}

Similar to Assumption \ref{assumption2}(i), Assumption \ref{assumption6}(i) imposes a condition on the second moment of $\widetilde{\bm{\mu}}^{(k)}$. The primary purpose of Assumption \ref{assumption6}(ii) is to ensure that $\mathbb{E}(\widetilde{\xi}^{-1} \sup_{\bm{\omega} \in \mathcal{W}} | \| \widehat{\bm{\mu}}(\bm{\omega})-\bm{\mu}^{(0)}\|^2 -\|\widetilde{\bm{\mu}}(\bm{\omega})-\bm{\mu}^{(0)}\|^2 | \mid \mathcal{X})=o_p(1)$, as $\widetilde{\xi}^{-1} \sup_{\bm{\omega} \in \mathcal{W}} | \| \widehat{\bm{\mu}}(\bm{\omega})-\bm{\mu}^{(0)}\|^2 -\|\widetilde{\bm{\mu}}(\bm{\omega})-\bm{\mu}^{(0)}\|^2 |=o_p(1)$ can be derived from Assumptions \ref{assumption2}(i), \ref{assumption4}, \ref{assumption5}, \ref{assumption6}(i) and \ref{assumption6}(iii) in Section \ref{proof2-2} of the supplementary material.

Assumption \ref{assumption6}(iii) imposes a constraint on the relation among $\widetilde{\xi}$,  $n_0$, $\bar{\alpha}$, and $K$. Similar constraints on $\widetilde{\xi}$ appear widely in the model averaging literature, including Eq. (15) of \citet{hansen2007least}, Eq. (8) of \citet{Wan2010}, Assumption 2 of \citet{b24}, Assumption 5 of \citet{zhang2023model}, Assumption 5 of \citet{zhang2024prediction},  Condition 8 of \citet{hu2023optimal}, and related works. This assumption provides three distinct insights. First, Assumption \ref{assumption6}(iii) requires $\widetilde{\xi} > 0$, which will be violated if the target model is correct or if $\widetilde{\bm{\mu}}^{(k)} = \bm{\mu}^{(0)}$ for any $k = 1, \dots, K$, since in such cases $\widetilde{\xi}=\inf_{ \bm{\omega} \in \mathcal{W} } \widetilde{R}(\bm{\omega})  =\inf_{ \bm{\omega} \in \mathcal{W} } \mathbb{E} ( \| \widetilde{\bm{\mu}}(\bm{\omega})-\bm{\mu}^{(0)}\|^2  \mid  \bm{X}^{(0)} )=0$.  Second, Assumption \ref{assumption6}(iii) requires that $\widetilde{\xi} \gg \bar{\alpha}^{1/2} Kn_0$. When $K=O(1)$, this reduces to $\widetilde{\xi} \gg \bar{\alpha}^{1/2} n_0$. In fact, as long as the target model is slightly misspecified and the source models differ modestly from the assumed target model, $\widetilde{\xi}$ can accumulate at the order of $n_0$ as $n_0$ grows.  Moreover, we have $\bar{\alpha}^{1/2}=o(1)$ according to Assumption~\ref{assumption4}. These together imply that the above condition is mild. Third, we have
\begin{align}\label{eq6}
    \widetilde{\xi} & =\inf_{\bm{\omega}\in\mathcal{W}}\widetilde{R}(\bm{\omega}) \le \sup_{k\in\mathcal{K}}\|\widetilde{\bm{\mu}}^{(k)}-\bm{\mu}^{(0)}\|^2
   \le \sum_{k\in\mathcal{K}}\|\widetilde{\bm{\mu}}^{(k)}-\bm{\mu}^{(0)}\|^2 \nonumber \\
   & \le \sum_{k\in\mathcal{K}} \bigl\{ 2 \|\widetilde{\bm{\mu}}^{(k)}\|^2+2\|\bm{\mu}^{(0)}\|^2 \bigr\} = O_p(Kn_0),
\end{align}
where the last step follows from Assumptions~\ref{assumption2}(i), \ref{assumption6}(i), and Markov's inequality. Together with this result and  $\bar{\alpha}^{1/2}=o(1)$ (as implied by Assumption~\ref{assumption4}), Assumption~\ref{assumption6}(iii) appears to be reasonable and feasible under typical misspecification settings. 

As discussed above, Assumption \ref{assumption6}(iii) requires the target model to be misspecified. Therefore, the large-sample property under a correctly specified target model is discussed in Section \ref{theo:weight}.

\begin{theorem}\label{theorem2}
If Assumptions \ref{assumption2}-\ref{assumption6} hold, then
\begin{align*}
\frac{R(\widehat{\bm{\omega}})}{\inf_{ \bm{\omega} \in \mathcal{W} }R({\bm{\omega}})} = 1+ O_p( \widetilde{\xi}^{-1} \bar{\alpha}^{1/2} Kn_0) = 1+o_p(1).
\end{align*}
\end{theorem}

Theorem \ref{theorem2} is a statement of asymptotic optimality regarding the expected squared prediction loss, which is also explored in \citet{JMA2012}. It establishes that the squared in-sample prediction risk of the GTLMMA prediction $\widehat{\bm{\mu}}(\widehat{\bm{\omega}})$ becomes asymptotically equivalent to the infeasible optimal squared in-sample prediction risk, when $\bm{\omega}$ is restricted to the weight set $\mathcal{W}$. Moreover, a larger $\widetilde{\xi}$, corresponding to a greater discrepancy between $\widetilde{\bm{\mu}}^{(k)}$ and $\bm{\mu}^{(0)}$, accelerates the convergence of the prediction risk to the minimal risk. Likewise, a smaller $\bar{\alpha}$, which implies a faster parameter estimation convergence rate, further facilitates this convergence. Furthermore, it underscores the effectiveness of GTLMMA in mitigating the risk of negative transfer, because its predictive performance is no worse than that of the target model estimated without transferring, i.e., using only the target data.

Next, we discuss the asymptotic optimality of GTLMMA about out-of-sample prediction. 
In this article, out-of-sample prediction refers to forecasting future values. Denote the new samples by $\bm{y}_{\mathrm{new}}^{(0)}$ and $\bm{X}_{\mathrm{new}}^{(0)}$. For simplicity, we assume that the new sample also has size $n_0$. We also assume that $\bm{X}_{\mathrm{new}}^{(0)}$ come from the target population and that the new samples $(\bm{y}_{\mathrm{new}}^{(0)},\bm{X}_{\mathrm{new}}^{(0)})$ follow the target model. Define $\bm{\mu}_{\mathrm{new}}^{(0)}=\mathbb{E}(\bm{y}_{\mathrm{new}}^{(0)} \mid \bm{X}_{\mathrm{new}}^{(0)})$. Our task is to predict $\bm{y}_{\mathrm{new}}^{(0)}$ or $\bm{\mu}_{\mathrm{new}}^{(0)}$ based on $\bm{X}^{(k)}$ and $\bm{y}^{(k)}$ for $k=0,\dots,K$, along with $\bm{X}_{\mathrm{new}}^{(0)}$.

Define the new predictions as $\widehat{\bm{\mu}}_{\mathrm{new}}^{(k)}=f(\bm{X}_{\mathrm{new}}^{(0)},\widehat{\bm{\delta}}^{(k)})$. For instance, in a linear regression model, we have $\widehat{\bm{\mu}}_{\mathrm{new}}^{(k)}=\bm{X}_{\mathrm{new}}^{(0)}\widehat{\bm{\delta}}^{(k)}$, whereas in a SAR model, $\widehat{\bm{\mu}}_{\mathrm{new}}^{(k)}=(\bm{I}_{n_0} -  \widehat{\rho}^{(k)}\bm{W}_0)^{-1}\bm{X}_{\mathrm{new}}^{(0)}\widehat{\bm{\beta}}^{(k)}$, with the spatial weight matrix $\bm{W}_0$ left unchanged since we are predicting future values in the same region. 
The aggregated prediction is then defined as $\widehat{\bm{\mu}}_{\mathrm{new}}(\bm{\omega})=\sum_{k=0}^K \omega_k \widehat{\bm{\mu}}_{\mathrm{new}}^{(k)}$, and the corresponding out-of-sample prediction risk is $R_{\mathrm{new}}(\bm{\omega})= \mathbb{E} ( \| \widehat{\bm{\mu}}_{\mathrm{new}}(\bm{\omega})-\bm{\mu}_{\mathrm{new}}^{(0)}\|^2 \mid \mathcal{X}, \bm{X}_{\mathrm{new}}^{(0)} )$. Similarly to the definitions under the in-sample asymptotic optimality, we define $\overline{\bm{\mu}}_{\mathrm{new}}^{(k)}=f(\bm{X}_{\mathrm{new}}^{(0)},\overline{\bm{\delta}}^{(k)})$ with $\overline{\bm{\delta}}^{(k)} \in \bm{\Delta}^{(k)}$, $\widetilde{\bm{\mu}}_{\mathrm{new}}^{(k)}=f(\bm{X}_{\mathrm{new}}^{(0)},\bm{\delta}^{(k)*})$, $\widetilde{\bm{\mu}}_{\mathrm{new}}(\bm{\omega})=\sum_{k=0}^K \omega_k \widetilde{\bm{\mu}}_{\mathrm{new}}^{(k)}$, the risk $\widetilde{R}_{\mathrm{new}}(\bm{\omega})=\mathbb{E} ( \| \widetilde{\bm{\mu}}_{\mathrm{new}}(\bm{\omega})-\bm{\mu}_{\mathrm{new}}^{(0)}\|^2  \mid  \bm{X}_{\mathrm{new}}^{(0)} )$, and $\widetilde{\xi}_{\mathrm{new}}=\inf_{ \bm{\omega} \in \mathcal{W} } \widetilde{R}_{\mathrm{new}}(\bm{\omega})$.  We further need the following assumptions.

\begin{assumption}\label{assumption7}
\leavevmode
\begin{enumerate}
    \item For $i=1,\dots,n_0$, let $\mu_{i}^{(0)}$, $\widetilde{\mu}_{i}^{(k)}$, $\mu_{\mathrm{new},i}^{(0)}$, and $\widetilde{\mu}_{\mathrm{new},i}^{(k)}$ denote the $i$th components of the corresponding mean vectors $\bm{\mu}^{(0)}$, $\widetilde{\bm{\mu}}^{(k)}$, $\bm{\mu}_{\mathrm{new}}^{(0)}$, and $\widetilde{\bm{\mu}}_{\mathrm{new}}^{(k)}$. We assume $\mathbb{E}(\mu_{i}^{(0)})^4 =O(1)$, $\mathbb{E}( \widetilde{\mu}_{i}^{(k)} )^4 =O(1)$,  $\mathbb{E}(\mu_{\mathrm{new},i}^{(0)})^4 =O(1)$, and $\mathbb{E}( \widetilde{\mu}_{\mathrm{new},i}^{(k)} )^4 =O(1)$ over $k \in \mathcal{K}$ and $i=1,\dots,n_0$.
    \item $\widetilde{\xi}_{\mathrm{new}}^{-1} \sup_{\bm{\omega} \in \mathcal{W}} | \| \widehat{\bm{\mu}}_{\mathrm{new}}(\bm{\omega})-\bm{\mu}_{\mathrm{new}}^{(0)}\|^2 -\|\widetilde{\bm{\mu}}_{\mathrm{new}}(\bm{\omega})-\bm{\mu}_{\mathrm{new}}^{(0)}\|^2 | $ is uniformly integrable. 
    \item $\widetilde{\xi}_{\mathrm{new}}^{-1} \{\bar{\alpha}^{1/2} \vee (Kn_0^{-1/2})\} K n_0=o(1)$ almost surely. 
\end{enumerate}
\end{assumption}

Assumptions \ref{assumption7}(i)-(ii) are mild constraints related to the boundedness of moments and integrability, similar to Assumption 2 of \citet{zhang2024prediction}. Assumption~\ref{assumption7}(iii) is similar to Assumption~\ref{assumption6}(iii) in that it also requires the target model to be misspecified. Specifically, if there exist new samples $(\bm{X}_{\mathrm{new}}^{(0)},\bm{y}_{\mathrm{new}}^{(0)})$ such that $\widetilde{\bm{\mu}}_{\mathrm{new}}^{(k)}=\bm{\mu}_{\mathrm{new}}^{(0)}$ for any $k=0,\dots,K$, then $\widetilde{\xi}_{\mathrm{new}}=\inf_{ \bm{\omega} \in \mathcal{W} } \widetilde{R}_{\mathrm{new}}(\bm{\omega}) =0$,  which would violate Assumption~\ref{assumption7}(iii). Compared with Assumption~\ref{assumption6}(iii), Assumption~\ref{assumption7}(iii) further requires that $K = o(n_0^{1/2})$, which follows from the requirement that $\widetilde{\xi}_{\mathrm{new}}$ must grow faster than $\bar{\alpha}^{1/2} K n_0$ and $K^2 n_0^{1/2}$, together with the bound $\widetilde{\xi}_{\mathrm{new}} = O_p(K n_0)$.

\begin{theorem}\label{theorem3}

If Assumption \ref{assumption5} holds for $\overline{\bm{\mu}}_{\mathrm{new}}^{(k)}$ (i.e., after replacing $\overline{\bm{\mu}}^{(k)}$), and Assumptions \ref{assumption2}-\ref{assumption4}, \ref{assumption6}(i) and \ref{assumption7} hold, then
\begin{align*}
\frac{R_{\mathrm{new}}(\widehat{\bm{\omega}})}{\inf_{ \bm{\omega} \in \mathcal{W} }R_{\mathrm{new}}({\bm{\omega}})} = 1+ O_p\left(\widetilde{\xi}_{\mathrm{new}}^{-1}  \{\bar{\alpha}^{1/2} \vee (Kn_0^{-1/2})\} K n_0\right) = 1+o_p(1).
\end{align*}
\end{theorem}

Compared with Theorem \ref{theorem2}, which establishes the asymptotic optimality of in-sample prediction risk, Theorem \ref{theorem3} addresses the out-of-sample case, offering insights that are particularly valuable for practical applications.

\subsection{Weight consistency}\label{theo:weight}
In this section, since Assumptions \ref{assumption6}(iii) and \ref{assumption7}(iii) restrict our optimality theorems to cases where the target model is misspecified, we provide a theoretical justification for GTLMMA under the correctly specified target model.
Specifically, we derive the convergence rate for the total weight assigned to the informative set.

First, we provide a mathematical definition of the informative set. For any non-negative constant $h$, we can divide $\mathcal{K}$ into two disjoint subsets, $\mathcal{K}_h$ and $\mathcal{K}_h^c$, where $\mathcal{K}_h=\{k: \| \bm{\delta}^{(k)*}-\bm{\delta}^{(0)}\|^2 \le h, k \in \mathcal{K}\}$ and $\mathcal{K}_h^c$ denotes the complement of $\mathcal{K}_h$. 
Based on Assumption~\ref{assumption4}, a smaller $h$ implies that the models contained within the set $\mathcal{K}_h$ are more suitable for parameter-transfer. In the asymptotic analysis, this is formalized by requiring $h=o(1)$, under which we refer to $\mathcal{K}_h$ as the informative set, and the corresponding source models in $\mathcal{K}_h$ as informative models. The correct target model is obviously included in $\mathcal{K}_h$ for any $h \ge 0$, because $\bm{\delta}^{(0)*}=\bm{\delta}^{(0)}$. Notably, the definition of the set $\mathcal{K}_h$ is analogous to the informative sets proposed by \citet{li2022transfer} and \citet{zeng2026transfer}. However, the key difference is that their approaches rely on comparing the true parameter values between the source and target models. As a result, models included in $\mathcal{K}_h$ are useful for transfer only when all models are correctly specified. In contrast, our method compares the pseudo-true values of the parameters with the true target parameter, so source models in $\mathcal{K}_h$ can be useful for transfer even if misspecified. Compared to \citet{hu2023optimal}, our definition of the set $\mathcal{K}_h$ is broadened from requiring $\bm{\delta}^{(k)*} = \bm{\delta}^{(0)}$ to $\| \bm{\delta}^{(k)*}-\bm{\delta}^{(0)}\|^2 \le h$ with a non-negative constant $h$. 

Denote the sum of the weight estimates assigned to the models in $\mathcal{K}_h$ by $\widehat{\nu}=\sum_{k \in \mathcal{K}_h} \widehat{\omega}_k$. Let $\mathcal{W}_C=\{\bm{\omega} \in \mathcal{W}: \sum_{k \in \mathcal{K}_h} \omega_k=0\}$ be the subset of $\mathcal{W}$ that assigns all weights to the models in $\mathcal{K}_h^c$. Let $\widetilde{\xi}_C=\inf_{\bm{\omega} \in \mathcal{W}_C} \widetilde{R}(\bm{\omega})$ denote the optimal risk when weights are not assigned to models in $\mathcal{K}_h$. The required assumption is as follows.

\begin{assumption}\label{assumption8} 
$\widetilde{\xi}_C^{-1} [n_0  \{ (K\bar{\alpha}^{1/2}) \vee h\}]=o_p(1)$.
\end{assumption}

Assumption \ref{assumption8} imposes rate conditions on $K$, $n_0$, $h$, $\bar{\alpha}$ and $\widetilde{\xi}_C$ to ensure that $\widehat{\nu}=1-o_p(1)$. It creates a clear distinction between the performance of the models in $\mathcal{K}_h$ and $\mathcal{K}_h^c$. Specifically, using any informative models with the correct target model yields $\widetilde{R}(\bm{\omega})=O_p( n_0h)$, where the specific derivation step can be found in Section \ref{proof4} of the supplementary material. In contrast, the risk $\widetilde{R}(\bm{\omega})$ is asymptotically much larger than the maximum of $n_0 \bar{\alpha}^{1/2}K$ and $n_0h$ when $\bm{\omega}$ assigns all weight to models in $\mathcal{K}_h^c$. Therefore, the models in $\mathcal{K}_h$ can be distinguished from those outside in terms of their contributions to the target prediction task.

In addition, by applying the same argument as in \eqref{eq6} for $\widetilde{\xi}$ and using Assumptions~\ref{assumption2}(i) and \ref{assumption6}(i), we obtain $\widetilde{\xi}_C = O_p(n_0 K)$. Therefore, Assumption~\ref{assumption8} requires that $\bar{\alpha}^{1/2} = o(1)$ and $h/K = o(1)$. The former condition follows from the rate condition associated with Assumption~\ref{assumption4}. Moreover, under the condition $h=o(1)$ defining the informative set, we have $h/K=o(1)$ automatically, indicating that Assumption~\ref{assumption8} is mild.

\begin{assumption}\label{assumption9}
    Define the local neighborhood of $\bm{\delta}^{(0)}$ by $\bm{\Gamma} = \left\{\overline{\bm{\delta}} : h^{-1/2} \|\overline{\bm{\delta}} - \bm{\delta}^{(0)} \| < c_2 \right\}$, where $c_2$ is a positive constant. Assume that $f(\bm{X}^{(0)},\overline{\bm{\delta}})$ for any $\overline{\bm{\delta}} \in \bm{\Gamma}$ is differentiable with respect to $\overline{\bm{\delta}}$. Moreover, $\sup_{\overline{\bm{\delta}} \in \bm{\Gamma}} \| \partial f(\bm{X}^{(0)},\overline{\bm{\delta}})/ \partial \overline{\bm{\delta}}^\top \|^2 = O_p(n_0)$.
\end{assumption}

Under the assumption that the target model is correctly specified, Assumption~\ref{assumption9} provides a unified framework for parametric models, allowing us to avoid formulating separate assumptions for each model specification. Since Assumption \ref{assumption9} is conceptually similar to Assumption \ref{assumption5}, its implications may be understood by referring to the discussion of Assumption \ref{assumption5} in \ref{dis:ass}.

\begin{theorem}\label{theorem4}
Suppose that Assumptions \ref{assumption2}-\ref{assumption5}, \ref{assumption6}(i), \ref{assumption8} and \ref{assumption9} hold, then
\begin{align*}
   \widehat{\nu}=1-O_p\left([\widetilde{\xi}_C^{-1}n_0  \{ (K\bar{\alpha}^{1/2}) \vee h\}]^{1/2} 
   \right)=1-o_p(1).
\end{align*}
\end{theorem}

Theorem \ref{theorem4} demonstrates that GTLMMA asymptotically allocates weights to both the target model and the informative source models, and facilitates parameter-transfer only among models that may contribute positively, thereby reducing the risk of negative transfer to a certain extent. Given that the models in $\mathcal{K}_h^c$ may substantially differ from the target model, our method inherently excludes them and adaptively adjusts the weights in a data-driven manner. 

Compared to the weight consistency theorems in existing literature on model averaging in transfer learning (e.g., \citet{hu2023optimal} and \citet{zhang2024prediction}), we provide explicit convergence rate at which the total weight assigned to the models in the informative set $\mathcal{K}_h$ converges to 1. If $\widetilde{\xi}_C$ is larger, which means that the risk function of the models in $\mathcal{K}_h^c$ is larger, then the models within $\mathcal{K}_h$ and $\mathcal{K}_h^c$ are better distinguished, leading to a faster convergence of $\widehat{\nu}$ to 1. We also observe that decreasing $\bar{\alpha}$ and $h$ result in faster convergence of $\widehat{\nu}$ to 1.  These conclusions align well with intuition.

\section{Simulation studies} \label{sim}

In this section, we conduct simulation studies under the SAR model framework to evaluate the performance of GTLMMA, following the settings of \citet{zeng2026transfer}. Specifically, we set $n_0=256$, $K=20$, and $n_1=\dots=n_K=100$. We also set $p=20$, $s=3$, $\rho^{(0)}=0.4$, and $\bm{\beta}^{(0)}=(\bm{1}_{s}^{\top},\bm{0}_{p-s}^{\top})^{\top}$, where $\bm{1}_{s}$ denotes a $s \times 1$  vector with all components being 1, and $\bm{0}_{p-s}$ represents a $(p-s) \times 1$ vector with all components being 0. Let $\mathcal{A}$ denote a subset of the indices of the source models, for example, $\{1, \dots, K_1\}$ with $K_1 \leq K$. Define $\mathcal{H}_k$ as a random subset of $\{1,\dots,p\}$ with $|\mathcal{H}_k|=H$ if $k \in \mathcal{A}$ or $|\mathcal{H}_k|=p/2$ if $k \notin \mathcal{A}$. Set $H=5$. For a given $\mathcal{A}$ and $k=1,\dots,K$, $j=1,\dots,p$, if $k \in \mathcal{A}$, set $\rho^{(k)}=\rho^{(0)}$ and $\beta_j^{(k)}=\beta_j^{(0)}-0.05 \times \mathds{1}(j \in \mathcal{H}_k)$; if $k \notin \mathcal{A}$, set $\rho^{(k)}=-\rho^{(0)}$ and $\beta_j^{(k)}=\beta_j^{(0)}-2 \times \mathds{1}(j \in \mathcal{H}_k)$. The above setup indicates that the set $\mathcal{A}$ contains the indices of source models whose parameters are similar to those of the target model. Denote the cardinality of $\mathcal{A}$ as $|\mathcal{A}|$ and let $|\mathcal{A}| \in \{0,5,10,15,20\}$. 

For $k \in \mathcal{K}$ and $i=1,\dots,n_k$, the covariates $\bm{x}_i^{(k)}$ are independently generated from a multivariate normal distribution $N_p(\bm{0}_p,\bm{\Sigma}^{(k)})$ with $\bm{\Sigma}^{(k)}=\bm{I}_p$, and random errors $\epsilon_i^{(k)}$ are generated independently from a normal distribution $N(0,1)$. Additionally, we generate a new data set $\bm{X}_{\mathrm{new}}^{(0)}$ of size $n_0$ from the target population. We sequentially place points indexed from 1 to $n_0$ on an $n_0^{1/2} \times n_0^{1/2}$ grid. Based on the positions of these indexed points, we record their left-right adjacency relationships and use this information to construct the spatial weight matrix $\bm{W}_0$. The spatial weight matrices $\bm{W}_k$ for $k=1,\dots,K$ are constructed in a similar way to $\bm{W}_0$, and all of them reflect the interactions between grid points that are adjacent in the horizontal direction. Then we row-normalize the spatial weight matrices $\bm{W}_k$ for $k=0,\dots,K$, such that the elements of each row of $\bm{W}_k$ sum to unity.

We consider three model specification scenarios: (i) All candidate models are correctly specified; (ii) The target model is correctly specified, while the source models are partially misspecified, meaning that odd-numbered source models are correctly specified, and even-numbered ones are misspecified; (iii) The target model is misspecified, and the source models are partially misspecified in the same manner as in scenario (ii). The first setting is designed to be consistent with the simulation design in \citet{zeng2026transfer}. The latter two settings are considered because model misspecification is a common issue in real-world applications. For example, the spatial weight matrix is often misspecified in spatial econometric problems. To generate data from a misspecified model, we consider the following process: $\bm{y}^{(k)}= \rho^{(k)} \bm{W}_k^* \bm{y}^{(k)} + \bm{X}^{(k)}\bm{\beta}^{(k)} + \bm{\epsilon}^{(k)}$, where $\bm{W}_k^*$ is constructed in a similar but different manner from $\bm{W}_k$, capturing the top-bottom interactions between adjacent grid points in the vertical direction.

We compare the proposed GTLMMA with four methods introduced by \citet{zeng2026transfer}: Target-SAR, [K]-TranSAR, TranSAR, and Oracle TranSAR. While Target-SAR uses only the target sample, the other three methods incorporate the target model, but differ in how they utilize source models. Specifically, TranSAR adopts a data-driven approach to select transferable sources, [K]-TranSAR uses all sources, and Oracle TranSAR employs those in the set $\mathcal{A}$. All of these comparison methods are based on 2SLS. For GTLMMA, the estimation is based on MLE or 2SLS, and the resulting methods are referred to as GTLMMA-SAR (MLE) and GTLMMA-SAR (2SLS), respectively. For $k=0,\dots,K$, the 2SLS approach employs instrumental variables constructed from the original regressors $\bm{X}^{(k)}$ and their spatially weighted transformations $\bm{W}_k \bm{X}^{(k)}$.

The target parameter estimate $\widetilde{\bm{\delta}}^{(0)}$ is defined as $\widetilde{\bm{\delta}}^{(0)}=\sum_{k=0}^K\widehat{\omega}_k\widehat{\bm{\delta}}^{(k)}$ for GTLMMA, and as $\widetilde{\bm{\delta}}^{(0)}=\{(\widetilde{\bm{\beta}}^{(0)})^{\top},\widetilde{\rho}^{(0)}\}^{\top}$ for the other methods. To evaluate the estimation performance of different methods, we calculate the mean squared error by $\mathrm{MSE}_{\bm{\delta}}= \|\widetilde{\bm{\delta}}^{(0)}-\bm{\delta}^{(0)} \|^2$. Define $\bm{\mu}_{\mathrm{new}}^{(0)}=  (\bm{I}_{n_0}-\rho^{(0)} \bm{W}_0)^{-1}\bm{X}_{\mathrm{new}}^{(0)} \bm{\beta}^{(0)}$ when the target model is correct and $\bm{\mu}_{\mathrm{new}}^{(0)}=  (\bm{I}_{n_0}-\rho^{(0)} \bm{W}_0^*)^{-1}\bm{X}_{\mathrm{new}}^{(0)} \bm{\beta}^{(0)}$  when the target model is misspecified. Define $\widehat{\bm{\mu}}_{\mathrm{new}}=\sum_{k=0}^K \widehat{\omega}_k (\bm{I}_{n_0}-\widehat{\rho}^{(k)}\bm{W}_0)^{-1}\bm{X}_{\mathrm{new}}^{(0)} \widehat{\bm{\beta}}^{(k)}$ for GTLMMA and $\widehat{\bm{\mu}}_{\mathrm{new}}=(\bm{I}_{n_0}-\widetilde{\rho}^{(0)}\bm{W}_0)^{-1}\bm{X}_{\mathrm{new}}^{(0)} \widetilde{\bm{\beta}}^{(0)}$ for other methods. Then, we evaluate the predictive performance through $\mathrm{MSE}_{\bm{\mu}}=n_0^{-1}\|\widehat{\bm{\mu}}_{\mathrm{new}}-\bm{\mu}_{\mathrm{new}}^{(0)} \|^2$.

\begin{figure}[ht]
 \centerline{\includegraphics[width=5in]{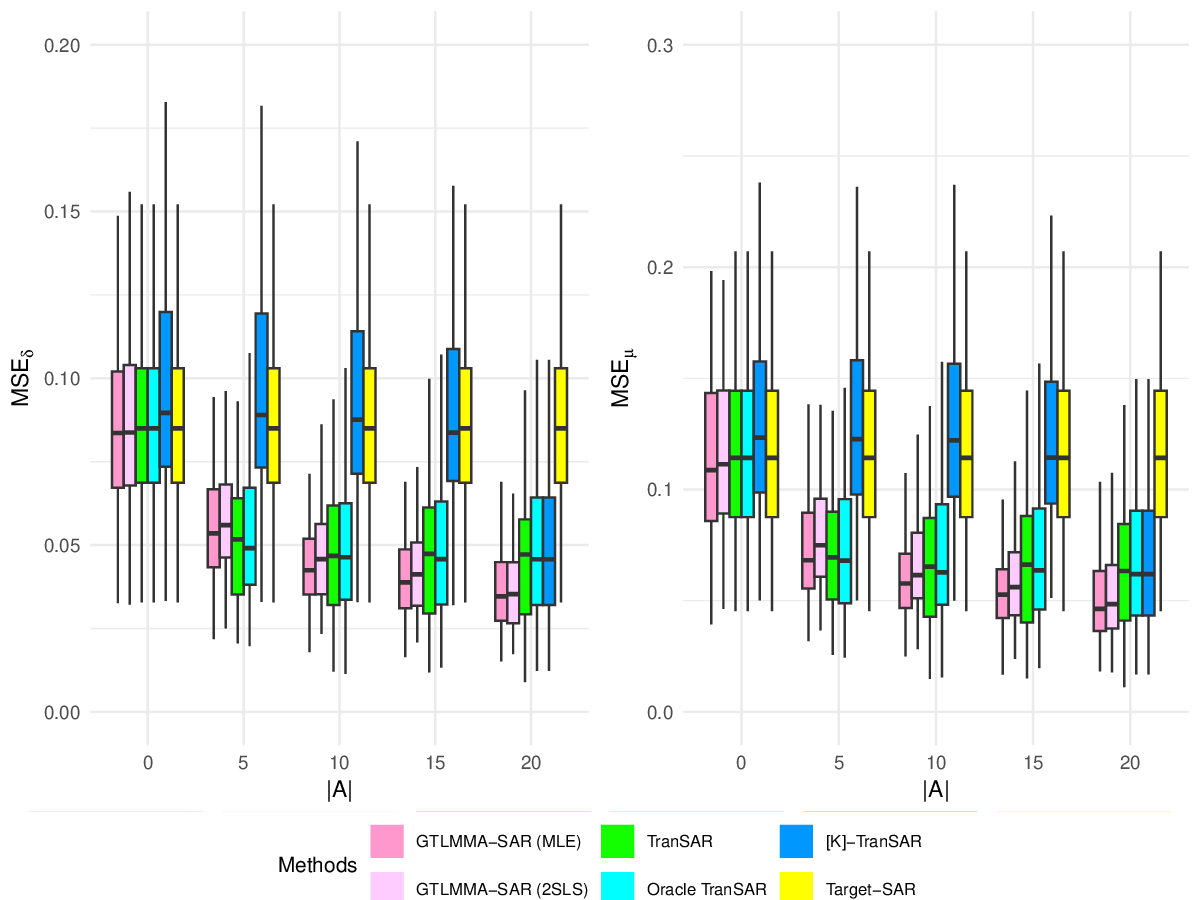}}
\caption{The boxplots of MSE$_{\bm{\delta}}$ and MSE$_{\bm{\mu}}$ of different methods when all candidate models are correct.}
\label{result0}
\end{figure}

\begin{figure}[ht]
 \centerline{\includegraphics[width=5in]{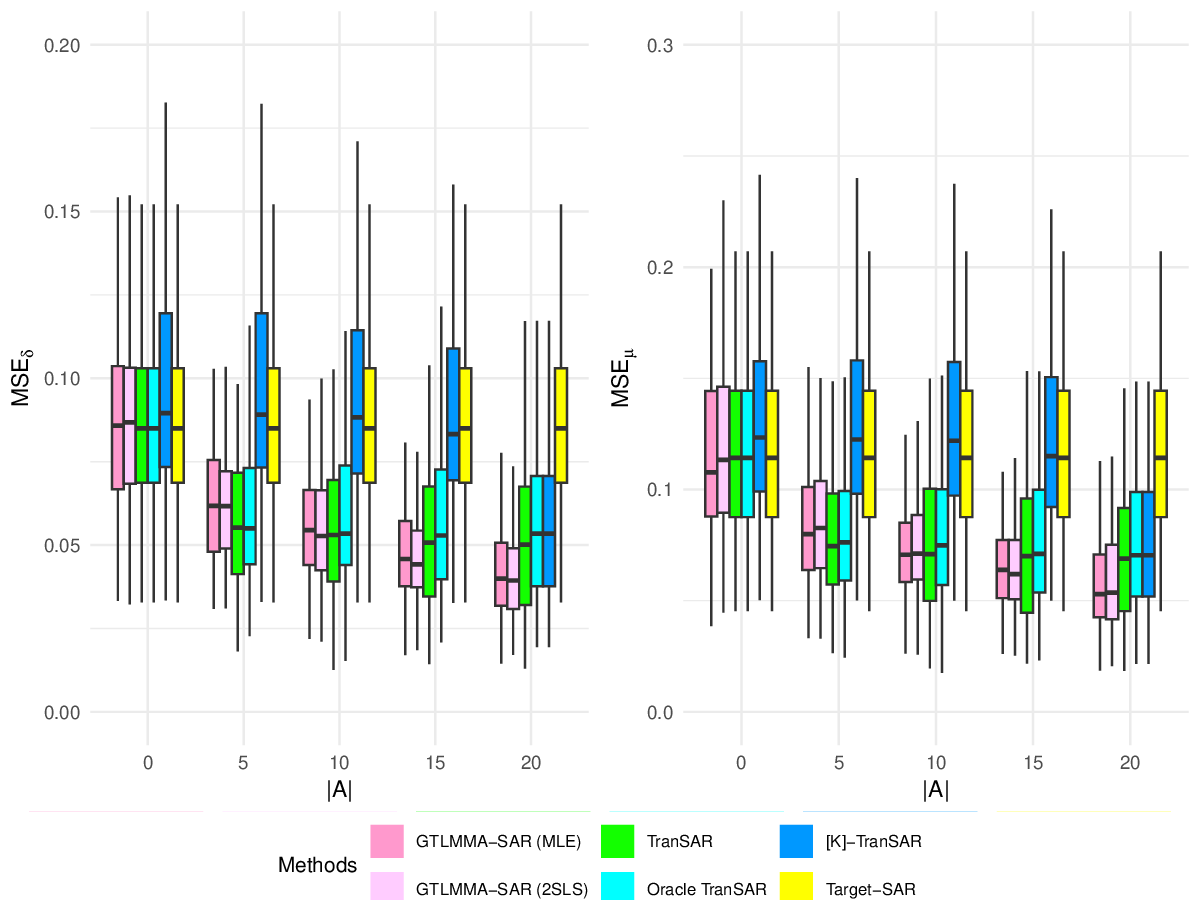}}
\caption{The boxplots of MSE$_{\bm{\delta}}$ and MSE$_{\bm{\mu}}$ of different methods with the correct target model and partially misspecified source models.}
\label{result1}
\end{figure}

\begin{figure}[ht]
 \centerline{\includegraphics[width=3.5in]{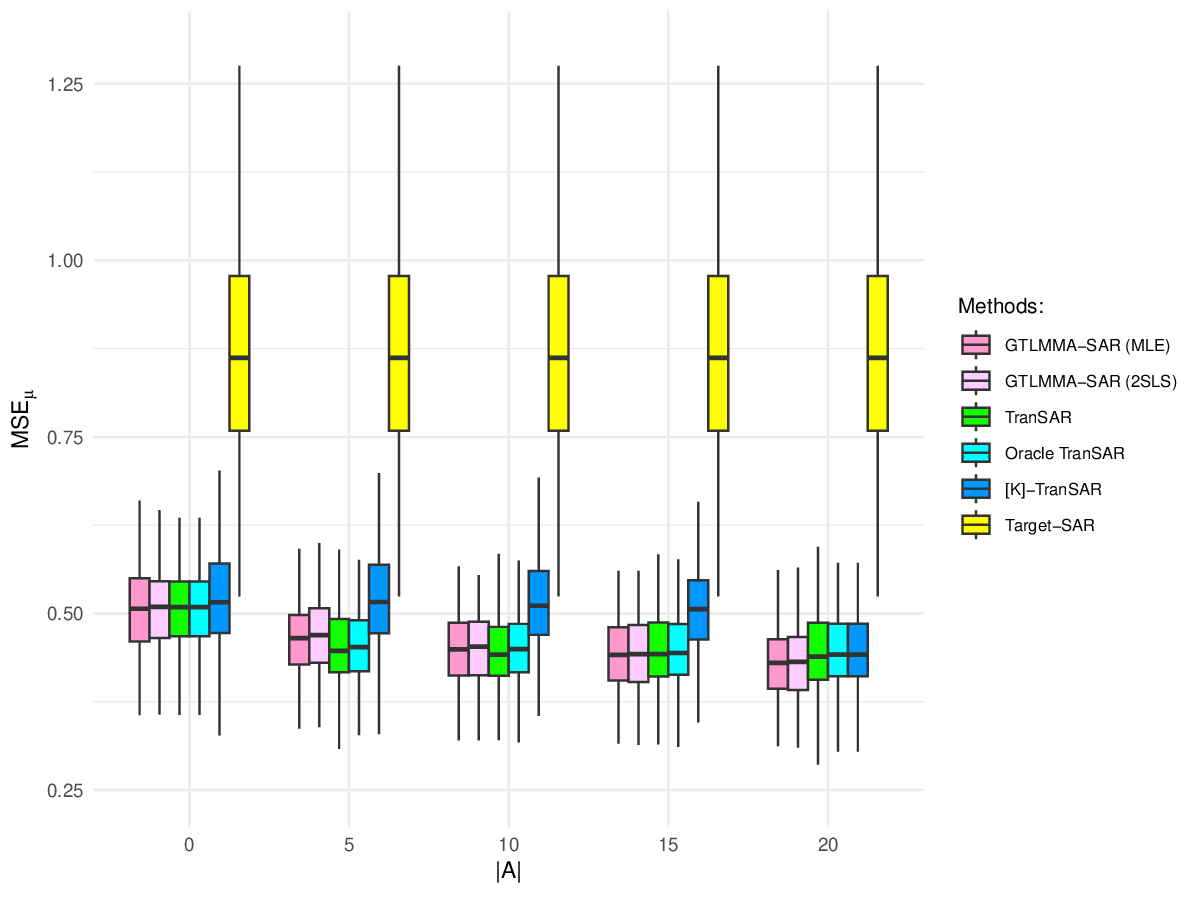}}
\caption{The boxplot of MSE$_{\bm{\mu}}$ of different methods with the misspecified target model and partially misspecified source models.}
\label{result2}
\end{figure}

Figures \ref{result0}-\ref{result2} present MSE$_{\bm{\delta}}$ and MSE$_{\bm{\mu}}$ based on 100 replications under different settings. The main conclusions are as follows. First, in all cases, GTLMMA-SAR (2SLS) improves over Target-SAR in terms of MSE$_{\bm{\mu}}$, demonstrating the effectiveness of the proposed parameter-transfer strategy. Overall, GTLMMA-SAR (MLE) and GTLMMA-SAR (2SLS) exhibit competitive performance and tend to perform relatively better as $|\mathcal{A}|$ increases. Second, when all candidate models are correctly specified, Oracle TranSAR outperforms TranSAR, and most transfer learning methods improve as $|\mathcal{A}|$ increases, with a more pronounced improvement observed for GTLMMA-SAR. Third, when the target model is correctly specified but the source models are partially misspecified, TranSAR tends to outperform Oracle TranSAR. This may be because the sources in $\mathcal{A}$ only share similar parameter values with the target, which does not necessarily transfer useful auxiliary information under misspecification. Fourth, when the target model is also misspecified, we only report MSE$_{\bm{\mu}}$ as measuring MSE$_{\bm{\delta}}$ becomes meaningless. In this case, GTLMMA-SAR (MLE) and GTLMMA-SAR (2SLS) tend to perform slightly better than other transfer methods when $|\mathcal{A}|=15$ and $20$.

Additionally, we analyze the relationship between $\widehat{\nu}$ (see Section \ref{theo:weight} for its definition) and the sample size when the target model is correctly specified. 
Specifically, we set $h=0$ and consider $K=6$ source models. All candidate models are correctly specified. In this setting, $\mathcal{K}_0$ (see Section~\ref{theo:weight}) denotes the set of informative source models. By construction, the first three source models belong to $\mathcal{K}_0$ and the remaining three belong to $\mathcal{K}_0^c$.

For $k=1,\dots,K$, if $k \in \mathcal{K}_0$, set $\bm{\delta}^{(k)}=\bm{\delta}^{(0)}$; if $k \notin \mathcal{K}_0$, set $\bm{\delta}^{(k)}=\bm{\delta}^{(0)}+c \bm{1}_{p+1}$, where $c=0.2$. We also set $n=n_0=n_1=\cdots=n_K$. Because the sample sizes determine the dimensions of the spatial weight matrices, and the SAR estimation procedures involve matrix determinant or matrix inversion operations, we focus on moderate sample sizes 
$n\in\{100,225,400,625,900,1024,1225\}$. All other settings are the same as in the previous simulations. For each sample size, we conduct 100 Monte Carlo replications.

\begin{figure}[ht]
 \centerline{\includegraphics[width=3.5in]{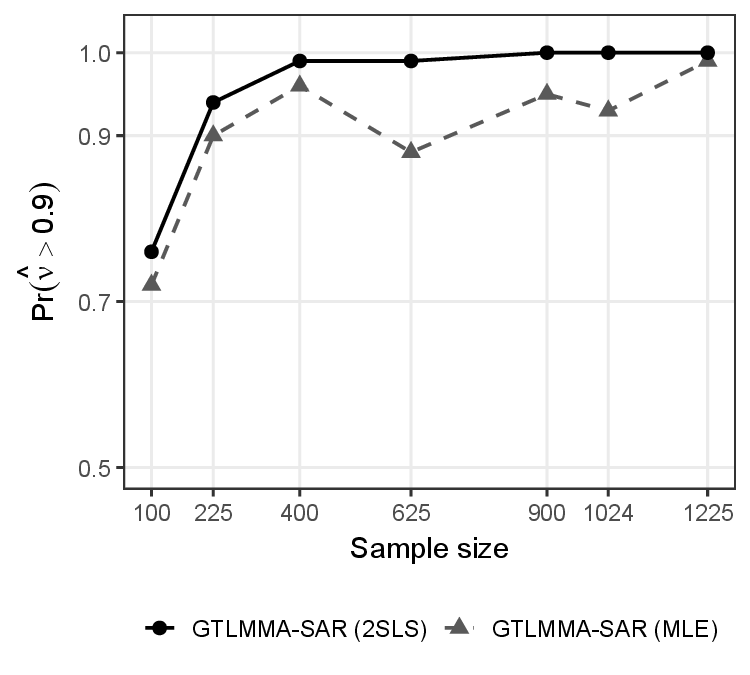}}
\caption{Empirical probability $\Pr(\widehat{\nu}>0.9)$ under different sample sizes.}
\label{fig-weight-consistency}
\end{figure}

To assess weight consistency empirically, Figure~\ref{fig-weight-consistency} reports the empirical probability $\Pr(\widehat{\nu}>0.9)$. 
The probability increases rapidly with the sample size for both GTLMMA-SAR implementations and becomes close to one when the sample size is moderately large, especially for the 2SLS implementation. 
The MLE implementation also assigns most of the total weight to the informative models, although it exhibits slightly larger finite-sample variability. These results provide finite-sample evidence supporting the weight consistency result in Theorem~\ref{theorem4}.

\section{Empirical application}\label{Empirical}

In this section, we apply the proposed method to predict monthly average house prices in selected U.S. metropolitan areas in 2023, using the \href{https://www.kaggle.com/datasets/shengkunwang/housets-dataset}{HouseTS} dataset. We focus on San Diego (SD), Los Angeles (LA), and San Francisco (SF), using monthly ZIP-code–level observations for each city. The monthly average house price is treated as the dependent variable. The covariates include inventory, median days on market, homes sold, total population, and pending sales, which are standard housing-market indicators capturing supply conditions, time-on-market dynamics, transaction volume, local market size, and near-term demand, respectively. These variables provide complementary information for house-price prediction.

To explicitly account for spatial dependence across neighboring ZIP codes, we adopt a SAR model. Spatial relationships are represented by ZIP-code–level spatial weight matrices $\bm{W}^{(k)}$, $k=0,1,2$, constructed separately for each city. For each ZIP code, neighbors are defined using a $k_{\mathrm{NN}}$-nearest-neighbor scheme with $k_{\mathrm{NN}}=5$,  based on ZIP Code Tabulation Area (ZCTA) geographic boundaries obtained from the \href{https://www.census.gov/geographies/mapping-files/time-series/geo/cartographic-boundary.2020.html#list-tab-1883739534}{U.S. Census cartographic boundary files}. All spatial weight matrices are row-normalized prior to estimation.

The goal is to improve predictive accuracy for the target city by transferring information on spatial dependence parameters $\rho^{(k)}$ and regression coefficients $\bm{\beta}^{(k)}$ from the source cities, thereby leveraging shared spatial patterns and common housing-market dynamics. To illustrate the applicability of our approach across different target tasks, we treat each city as the target in turn, with the remaining cities serving as sources. We implement the proposed GTLMMA-SAR (MLE) and GTLMMA-SAR (2SLS) methods, and other competitive methods, including TranSAR, [K]-TranSAR, and Target-SAR, to predict house prices in the target city. We do not consider Oracle TranSAR, because it is infeasible in applications owing to unknown information about auxiliary models.

Prediction is conducted using a rolling one-step-ahead forecasting scheme, where data from one month are used for training and house prices in the subsequent month are predicted. Let $\bm{X}_{\mathrm{test}} \in \mathbb{R}^{n_0 \times p}$ and $\bm{y}_{\mathrm{test}} \in \mathbb{R}^{n_0}$ denote the design matrix and the response vector of the test data, respectively. The test sample size equals the training sample size for each city, since the set of ZIP codes remains unchanged across months. For the proposed GTLMMA-SAR method, the predicted values for the test samples are given by $\widehat{\bm{y}}_{\mathrm{test}}=\sum_{k=0}^K \widehat{\omega}_k (\bm{I}_{n_0}-\widehat{\rho}^{(k)}\bm{W}_0)^{-1}\bm{X}_{\mathrm{test}} \widehat{\bm{\beta}}^{(k)}$, while for the competing methods, the prediction is given by $\widehat{\bm{y}}_{\mathrm{test}}=(\bm{I}_{n_0} - \widehat{\rho}\bm{W}_0)^{-1} \bm{X}_{\mathrm{test}}\widehat{\bm{\beta}}$. For prediction, all candidate models are evaluated using the target spatial weight matrix $\bm{W}_0$, and $\widehat{\rho}$, $\widehat{\rho}^{(k)}$, $\widehat{\bm{\beta}}$, and $\widehat{\bm{\beta}}^{(k)}$ denote the parameter estimates obtained using the training samples. Prediction accuracy for each forecasting step is evaluated using the mean squared prediction error (MSPE), $\mathrm{MSPE} = n_{\mathrm{test}}^{-1} \| \bm{y}_{\mathrm{test}}- \widehat{\bm{y}}_{\mathrm{test}} \|^2$. The rolling one-step-ahead forecasting scheme is repeated sequentially over the year, yielding 11 forecasts over the 12-month period, and the reported MSPE is obtained by averaging the corresponding prediction errors, with results for all methods reported in Table~\ref{count-result}.

Table~\ref{count-result} reports the MSPE for different methods. The GTLMMA-SAR estimator based on MLE attains the lowest MSPE across all three target cities. 
For the 2SLS-based implementations, GTLMMA-SAR attains the best result in Los Angeles and shows competitive performance in the other cases. In addition, transfer learning methods generally tend to outperform the target-only benchmark, indicating the benefit of incorporating information from related cities.

\begin{table}[ht] 
  \caption{The MSPE of different methods in HouseTS U.S. house-price prediction}
  \small
\begin{tabular}{cccccc}
\toprule
\multirow{3}{*}{\diagbox{Target}{Method}} & MLE-based & \multicolumn{4}{c}{2SLS-based} \\
 \cmidrule(lr){2-2} \cmidrule(lr){3-6} 
& \thead{GTLMMA-SAR\\(MLE)} & \thead{GTLMMA-SAR\\(2SLS)} & TranSAR & [K]-TranSAR & Target-SAR \\
\midrule
San Diego (SD) & \textbf{0.8089} & 1.0250 & 1.0088 & \emph{0.8273} & 1.0350   \\ 
Los Angeles (LA)  & \textbf{0.6703} & \emph{0.7259} & 0.8331 & 0.8403 & 0.9093  \\ 
San Francisco (SF) & \textbf{0.9493} & 0.9875 & 1.0433 & 1.0606 & \emph{0.9839}  \\ 
\bottomrule
\end{tabular}
 \label{count-result}
 \begin{tablenotes} 
 \item \textit{Note}: The \textbf{bold} results are the minimal values in each row, and the \emph{italic} results are the second minimal values in each row.
     \end{tablenotes} 
\end{table}

\section{Extensions}\label{exten}

In this section, we extend the proposed GTLMMA framework in three directions. First, we generalize the parametric framework to a semiparametric setting. Second, we extend it from cross-sectional data to panel data through a linear mixed-effects model. Finally, we discuss a nonparametric extension to further broaden its applicability.

\subsection{Semiparametric extension}

    We consider an additive partially linear model \citep{Engle1986, Speckman1988}. For $k=0,\dots,K$, the working mean is $\bm{\mu}^{(k)}_{\mathrm{wm}} = \bm{Z}^{(k)}\bm{\delta}^{(k)} + \bm{g}^{(k)}(\bm{V}^{(k)})$, where $\bm{Z}^{(k)} \in \mathbb{R}^{n_k \times p}$ enters the parametric component and $\bm{V}^{(k)} \in \mathbb{R}^{n_k \times q_k}$ (with $q_k = p_k - p$) enters the additive nonparametric component. In the semiparametric setting, we transfer only the parametric component across datasets, while the nonparametric component is estimated from the target data. Technical details are provided in \ref{calcu:semi}.

    To construct the GTLMMA criterion, we approximate the target nonparametric component using spline bases, which yields a linear representation $\bm{F}^{(0)} = [\bm{Z}^{(0)}, \bm{B}^{(0)}]$, where $\bm{B}^{(0)}$ is the spline design matrix for the target nonparametric component. Let $t_l^{(0)}$ denote the number of spline basis functions used to approximate the $l$th component, and define $r_0 = p + \sum_{l=1}^{q_0} t_l^{(0)}$. Let $ \widehat{\bm{\theta}}^{(0)} = \{(\bm{F}^{(0)})^{\top}\bm{F}^{(0)}\}^{-1} (\bm{F}^{(0)})^{\top}\bm{y}^{(0)} = \{(\widehat{\bm{\delta}}^{(0)})^{\top},(\widehat{\bm{\gamma}}^{(0)})^{\top}\}^{\top}$. Define the selection matrices $\bm{S}_{\delta} = (\bm{I}_p, \bm{0})$ and $\bm{S}_{\gamma} = (\bm{0}, \bm{I}_{r_0 - p})$. To characterize the derivative structure, define $\bm{H}_{\delta}^{(0)}=\bm{Z}^{(0)}\bm{S}_{\delta}\{(\bm{F}^{(0)})^{\top}\bm{F}^{(0)}\}^{-1}(\bm{F}^{(0)})^{\top}$ and $\bm{H}_{\gamma}^{(0)}=\bm{B}^{(0)}\bm{S}_{\gamma}\{(\bm{F}^{(0)})^{\top}\bm{F}^{(0)}\}^{-1}(\bm{F}^{(0)})^{\top}$. Then $\partial \widehat{\bm{\mu}}^{(0)} / \partial(\bm{y}^{(0)})^{\top}= \bm{H}_{\delta}^{(0)} + \bm{H}_{\gamma}^{(0)}$. For $k=1,\dots,K$, the transfer predictor is $\widehat{\bm{\mu}}^{(k)}=\bm{Z}^{(0)}\widehat{\bm{\delta}}^{(k)}+\bm{B}^{(0)}\widehat{\bm{\gamma}}^{(0)}$, which implies $\partial \widehat{\bm{\mu}}^{(k)}/\partial(\bm{y}^{(0)})^{\top}=\bm{H}_{\gamma}^{(0)}$. Therefore, $\partial \widehat{\bm{\mu}}(\bm{\omega}) / \partial(\bm{y}^{(0)})^{\top}=\omega_0 \bm{H}_{\delta}^{(0)} + \bm{H}_{\gamma}^{(0)}$. Based on this structure, the Mallows-type criterion takes the form
    \begin{align*}
    \mathcal{C}_{\mathrm{semi}}(\bm{\omega}) = \|\bm{y}^{(0)} - \widehat{\bm{\mu}}(\bm{\omega})\|^2 + 2 \omega_0 \mathrm{tr}\{\bm{H}_{\delta}^{(0)} \bm{\Omega}_0\}.
    \end{align*}

    The following result shows that $\mathcal{C}_{\mathrm{semi}}(\bm{\omega})$ is conditionally unbiased for the prediction risk up to an additive term. The additional term $- 2 \mathrm{tr}\{\bm{H}_{\gamma}^{(0)} \bm{\Omega}_0\}$ arises from the nonparametric component, which does not depend on $\bm{\omega}$.

    \begin{proposition}\label{proposition:semi:unbiased}
    Under the semiparametric model, $\mathbb{E}\{\mathcal{C}_{\mathrm{semi}}(\bm{\omega}) \mid \mathcal{X}\} =R(\bm{\omega}) + \mathrm{tr}(\bm{\Omega}_0) - 2 \mathrm{tr}\{\bm{H}_{\gamma}^{(0)} \bm{\Omega}_0\}$.
    \end{proposition}    

    Based on this criterion, we define the weights by minimizing its empirical version. Specifically, we define $\widehat{\bm{\omega}}=\arg\min_{\bm{\omega} \in \mathcal{W}}\widehat{\mathcal{C}}_{\mathrm{semi}}(\bm{\omega})$, where $\widehat{\mathcal{C}}_{\mathrm{semi}}(\bm{\omega})$ is obtained from $\mathcal{C}_{\mathrm{semi}}(\bm{\omega})$ by replacing $\bm{\Omega}_0$ with its estimator $\widehat{\bm{\Omega}}_0$.

    Compared with the parametric specification, the semiparametric model introduces an additional nonparametric component, which requires modified rate conditions. For $k=0,\dots,K$, define $r_k=p + \sum_{l=1}^{q_k} t_l^{(k)}$, $\alpha_{k,\mathrm{semi}} =p/n_k +(r_0-p) / n_0$, and $\bar{\alpha}_{\mathrm{semi}}=\sup_{k \in \mathcal{K}} \alpha_{k,\mathrm{semi}}$. The detailed assumptions and notation are given in \ref{sec:ass:semi}. In what follows, $R(\bm{\omega})$ and $R_{\mathrm{new}}(\bm{\omega})$ are defined as in the parametric setting and evaluated under the semiparametric model. The following proposition establishes the corresponding asymptotic optimality results for the semiparametric criterion.

   \begin{proposition}\label{proposition:semi:optimality}
   \textbf{(i) In-sample optimality.}
   Suppose that Assumptions~\ref{assumption2}, \ref{assumption4}, \ref{assumption5}, \ref{assumption6} and \ref{ass:semi:eig} hold under the semiparametric notation in Table~\ref{semi:notation}. Then,
   \[
   \frac{R(\widehat{\bm{\omega}})}{\inf_{\bm{\omega}\in\mathcal{W}}R(\bm{\omega})}=1+O_p\left(\widetilde{\xi}^{-1}\bar{\alpha}_{\mathrm{semi}}^{1/2}K n_0\right)=1+o_p(1).
   \]
   
   \textbf{(ii) Out-of-sample optimality.}
   Suppose that Assumption~\ref{assumption5} holds with $\overline{\bm{\mu}}^{(k)}$ replaced by $\overline{\bm{\mu}}_{\mathrm{new}}^{(k)}$, and that Assumptions~\ref{assumption2}, \ref{assumption4}, \ref{assumption6}(i), \ref{assumption7} and \ref{ass:semi:eig} hold under the semiparametric notation in Table~\ref{semi:notation}. Then,
   \[
   \frac{R_{\mathrm{new}}(\widehat{\bm{\omega}})}{\inf_{\bm{\omega}\in\mathcal{W}}R_{\mathrm{new}}(\bm{\omega})} = 1+O_p\left(\widetilde{\xi}_{\mathrm{new}}^{-1}\left\{\bar{\alpha}_{\mathrm{semi}}^{1/2}\vee (K n_0^{-1/2}) \right\}K n_0\right)=1+o_p(1).
   \]
   \end{proposition}

    The weight consistency result requires additional conditions beyond those used in the parametric setting. In particular, Assumption~\ref{assumption9} in Theorem~\ref{theorem4} does not extend to the semiparametric framework, as it is formulated for the parametric component only. We therefore introduce Assumption~\ref{assumption10}; see \ref{sec:ass:semi} for details. The corresponding result is given below.
    
    \begin{proposition}\label{proposition:semi:weight}
    Suppose that Assumptions \ref{assumption2}-\ref{assumption5}, \ref{assumption6}(i), \ref{assumption8} and \ref{assumption10} hold (under the semiparametric notation in Table~\ref{semi:notation}). Then
    \begin{align*}
    \widehat{\nu}=1-O_p\left([\widetilde{\xi}_C^{-1}  n_0  \{ (K\bar{\alpha}_{\mathrm{semi}}^{1/2}) \vee h\} ]^{1/2} 
   \right)=1-o_p(1).
   \end{align*}
   \end{proposition}
   Proposition~\ref{proposition:semi:weight} extends Theorem~\ref{theorem4} to the semiparametric setting. It shows that the estimated weights concentrate on the informative sources, with convergence rate governed by $\bar{\alpha}_{\mathrm{semi}}$. To illustrate this extension, an empirical example is provided in the supplementary material.

\subsection{Panel data extension} \label{exten:panel}

In this section, we extend GTLMMA from the cross-sectional setting to the panel data framework. As an illustration, we consider a linear mixed-effects model. For clarity, we defer the full model specification and explicit formulae for the panel GTLMMA criterion to \ref{calcu:panel}. 
We outline the key elements of the extension here.

For $k=0,\dots,K$, the $k$th dataset consists of $n_k$ clusters, where the $i$th cluster contains $T_i^{(k)}$ observations. We adopt a linear mixed-effects working model with working mean $\bm{\mu}^{(k)}_{\mathrm{wm}} = \bm{X}_{\mathrm{FE}}^{(k)} \bm{\delta}^{(k)} + \bm{Z}_{\mathrm{RE}}^{(k)} \bm{\vartheta}^{(k)}$, where $\bm{\delta}^{(k)}$ and $\bm{\vartheta}^{(k)}$ denote the fixed effects vector and the cluster-specific random effects, respectively. We define the target mean as $\bm{\mu}^{(0)}=\mathbb{E}(\bm{y}^{(0)} \mid \bm{X}_{\mathrm{FE}}^{(0)},\bm{Z}_{\mathrm{RE}}^{(0)},\bm{\vartheta}^{(0)})$, corresponding to a cluster-specific prediction setting where future observations arise from the same clusters and we condition on the random effects (see \citet{donohue2011conditional,zhang2014biometrika}). We transfer only the fixed effects $\bm{\delta}^{(k)}$ from the source models, whereas the random effects $\bm{\vartheta}^{(0)}$ and within-cluster variance are always estimated from the target data. 

The panel GTLMMA criterion follows the same construction principle as \eqref{eq4}, with $\widehat{\bm{\mu}}^{(0)}$ defined under the linear mixed-effects model. Let $\mathcal{X}_{\mathrm{panel}}=\{\bm{X}_{\mathrm{FE}}^{(0)},\dots,\bm{X}_{\mathrm{FE}}^{(K)},\bm{Z}_{\mathrm{RE}}^{(0)},\dots,\bm{Z}_{\mathrm{RE}}^{(K)},\\ \bm{\vartheta}^{(0)},\dots,\bm{\vartheta}^{(K)}\}$, and define the conditional prediction risk $R_{\mathrm{panel}}(\bm{\omega})=\mathbb{E}( \| \widehat{\bm{\mu}}_{\mathrm{panel}}(\bm{\omega}) - \bm{\mu}^{(0)} \|^2 \mid \mathcal{X}_{\mathrm{panel}})$. We next state a formal result establishing the conditional unbiasedness of the panel GTLMMA criterion.

\begin{proposition}\label{proposition:panel}
Under the linear mixed-effects specification described in \ref{calcu:panel}, the panel GTLMMA criterion \eqref{panel_cri1} is a conditionally unbiased estimator of the prediction risk: $\mathbb{E}\{\mathcal{C}_{\mathrm{panel}}(\bm{\omega}) \mid \mathcal{X}_{\mathrm{panel}}\}= R_{\mathrm{panel}}(\bm{\omega}) + \sigma_0^2 N_0$, where $N_0 = \sum_{i=1}^{n_0} T_i^{(0)}$.
\end{proposition}

\subsection{Nonparametric extension}\label{exten:non}

    We further consider a nonparametric setting in which the population-level conditional mean is treated as an unknown regression function. Specifically, the data generating process is written as $y_i^{(k)} = \mathbb{E}\{y_i^{(k)} \mid \widetilde{\bm{x}}_i^{(k)}\} + e_i^{(k)}$, where $\widetilde{\bm{x}}_i^{(k)} = (x_{i1}^{(k)}, x_{i2}^{(k)}, \ldots)^\top$ denotes a potentially infinite-dimensional covariate vector.
    

    To illustrate the extension of GTLMMA, we use random forests as a representative example. For $k=0,\dots,K$, a random forest predictor is constructed using $(\bm{X}^{(k)}, \bm{y}^{(k)})$ and evaluated at the target covariates $\bm{X}^{(0)}$. 
    For a fitted forest, the resulting predictor admits the linear representation $\widehat{\bm{\mu}}^{(k)}=\bm{P}_{k,\pi}^{(0)}\bm{y}^{(k)}$, where $\bm{P}_{k,\pi}^{(0)}\in\mathbb{R}^{n_0\times n_k}$ is the data-dependent weight matrix induced by the fitted forest. Detailed construction of this representation is provided in \ref{calcu:nonparametric}.

    Due to the highly complex relationship between $\widehat{\bm{\mu}}^{(0)}$ and $\bm{y}^{(0)}$ in random forest models, the exact form of $\partial \widehat{\bm{\mu}}^{(0)} / \partial (\bm{y}^{(0)})^{\top}$ is generally intractable. One possible approach is to employ the approximation method described in \ref{cau:derivative}. Alternatively, motivated by \citet{chen2024optimal}, we express the penalty term using the matrix $\bm{P}_{0,\pi}^{(0)}$. This matrix serves as a surrogate for the derivative $\partial \widehat{\bm{\mu}}^{(0)} / \partial (\bm{y}^{(0)})^{\top}$, thereby linking this formulation to the general criterion in \eqref{eq4}. We therefore define the GTLMMA weight selection criterion for random forests as
    \begin{align*}
        \widehat{\mathcal{C}}_{\mathrm{forest}}(\bm{\omega})= \|  \bm{y}^{(0)} - \widehat{\bm{\mu}}(\bm{\omega}) \|^2 + 2 \omega_0 \mathrm{tr}\left(\bm{P}_{0,\pi}^{(0)}\widehat{\bm{\Omega}}_0 \right).
    \end{align*}
    The optimal weight is then given by $\widehat{\bm{\omega}}=\underset{\bm{\omega} \in \mathcal{W} }{\arg\min} \,\widehat{\mathcal{C}}_{\mathrm{forest}}(\bm{\omega})$.

    This example illustrates how GTLMMA can be adapted to flexible nonparametric estimators through suitable prediction representations. Further implementation details and broader discussion of nonparametric extensions are provided in \ref{calcu:nonparametric}.

\section{Discussion} \label{Discussion}
In this paper, we introduce a generalized parameter-transfer approach based on the optimal Mallows-type model averaging, which transfers information from related domains through Mallows model averaging without requiring shared multi-source data. There are some relevant research topics worth exploring. First, our approach allows $p$ to diverge with $\underline{n}$ under large-sample properties but requires $p < \underline{n}$. Therefore, proposing a model averaging approach tailored to parameter-transfer under $p \ge \underline{n}$ is a meaningful direction for further exploration. Second, extending the GTLMMA method to generalized linear models would involve replacing the squared loss with the negative log-likelihood or Kullback–Leibler loss, and \citet{zhang2016optimal} may serve as a helpful reference for such an extension. 

\section*{Acknowledgement}
The authors are grateful to Hao Zeng for providing the code and valuable suggestions.

\section*{Competing interests}
The authors declare none.

\begin{center}
{\Large \bfseries APPENDIX}
\end{center}

 \appendix
\renewcommand{\thetable}{\Alph{section}.\arabic{table}}
\setcounter{table}{0}
\setcounter{assumption}{0}
\renewcommand{\theassumption}{\Alph{section}.\arabic{assumption}}
\setcounter{example}{0}
\renewcommand{\theexample}{\Alph{section}.\arabic{example}}

\section{Computing the derivative of fitted predictions}\label{cau:derivative}

In this section, we discuss the matrix $\partial \widehat{\bm{\mu}}^{(0)} / \partial (\bm{y}^{(0)})^{\top}$. Depending on the structure of the predictor, this derivative can be obtained in several ways.

\begin{enumerate}
    \item \textbf{Linear smoothers:} 
    For estimators admitting a linear representation $\widehat{\bm{\mu}}^{(0)} = \bm{Q}_0 \bm{y}^{(0)}$, where the smoothing matrix $\bm{Q}_0$ depends only on $\bm{X}^{(0)}$, the derivative is given by $\partial \widehat{\bm{\mu}}^{(0)} / \partial (\bm{y}^{(0)})^{\top} = \bm{Q}_0$. Examples include ordinary least squares, ridge regression, smoothing splines, and certain kernel methods \citep{hastie2009elements}.

    \item \textbf{Estimators defined through parameters:} 
    When the predictor depends on an unknown parameter $\bm{\delta}^{(0)}$, the fitted values take the form $\widehat{\bm{\mu}}^{(0)} = g(\bm{X}^{(0)}, \widehat{\bm{\delta}}^{(0)})$, where $\widehat{\bm{\delta}}^{(0)}$ is obtained from an optimization or estimating equation. The derivative with respect to $\bm{y}^{(0)}$ can then be derived via the implicit function theorem under standard regularity conditions. This applies to a broad class of smooth M-estimators, including Huber-type regression with differentiable approximations, ridge-type penalized regression, and maximum likelihood estimators under standard regularity conditions \citep{huber1992robust,newey1994large}. For differentiable parametric models, this derivative-based characterization is closely related to influence function arguments, which provide a first-order approximation by viewing the estimator as a functional of the empirical distribution \citep{hampel1974influence}.

    \item \textbf{Black-box predictors:}  For highly flexible machine learning algorithms, such as random forests or other ensemble methods, the explicit form of $\widehat{\bm{\mu}}^{(0)}$ is typically unavailable, making influence function-based calculations difficult to apply directly. In such cases, the infinitesimal jackknife provides a practical approximation based on perturbations of observation weights \citep{efron2014estimation,wager2014confidence}. Related approximations based on gradients and Hessians have also been proposed in modern machine learning settings \citep{koh2017understanding}.

\end{enumerate}

\section{Technical details for model specification and implementation}
This section provides model-specific technical details required to implement the proposed method under different model specifications.

\subsection{Nonlinear regression model} \label{calcu:nonli}

In this section, we show how to compute the diagonal elements of $\partial \widehat{\bm{\mu}}^{(0)}/\partial (\bm{y}^{(0)})^{\top}$ for nonlinear regression models, which are required for constructing the GTLMMA criterion.

We consider the nonlinear least-squares estimator obtained by minimizing
\[
L(\bm{\delta}^{(k)}) = n_k^{-1} \sum_{i=1}^{n_k} \big[ y_i^{(k)} - \phi(\bm{x}_i^{(k)}, \bm{\delta}^{(k)}) \big]^2
\]
over $\Theta_k$. For the target sample, define
\[
\bm{g}(\bm{\delta},\bm{y}^{(0)})=\sum_{i=1}^{n_0}\big[ y_i^{(0)}-\phi(\bm{x}_i^{(0)},\bm{\delta}) \big]\frac{\partial \phi(\bm{x}_i^{(0)},\bm{\delta})}{\partial \bm{\delta}}.
\]
Under standard regularity conditions (e.g., \citet{Jennrich1969}), the estimator $\widehat{\bm{\delta}}^{(0)}$ is well-defined and satisfies the first-order condition $\bm{g}(\widehat{\bm{\delta}}^{(0)},\bm{y}^{(0)})=\bm{0}$. Differentiating this condition with respect to $y_t^{(0)}$ yields
\begin{align}\label{eq5}
    \frac{\partial \bm{g}(\widehat{\bm{\delta}}^{(0)}, \bm{y}^{(0)})}{\partial y_t^{(0)}} + \frac{\partial \bm{g}(\widehat{\bm{\delta}}^{(0)}, \bm{y}^{(0)})}{\partial (\widehat{\bm{\delta}}^{(0)})^{\top}} \cdot \frac{\partial \widehat{\bm{\delta}}^{(0)}}{\partial y_t^{(0)}} = \bm{0}.
\end{align}
It follows from \eqref{eq5} that
\begin{align*}
    \frac{\partial \widehat{\mu}_t^{(0)}}{\partial y_t^{(0)}} & = \frac{\partial \phi(\bm{x}_t^{(0)}, \widehat{\bm{\delta}}^{(0)})}{\partial (\widehat{\bm{\delta}}^{(0)})^{\top}} \frac{\partial \widehat{\bm{\delta}}^{(0)}}{\partial y_t^{(0)}} =  - \frac{\partial \phi(\bm{x}_t^{(0)},\widehat{\bm{\delta}}^{(0)})}{\partial (\widehat{\bm{\delta}}^{(0)})^{\top}} \left\{\frac{\partial \bm{g}(\widehat{\bm{\delta}}^{(0)}, \bm{y}^{(0)})}{\partial (\widehat{\bm{\delta}}^{(0)})^{\top}}\right\}^{-1} \frac{\partial \bm{g}(\widehat{\bm{\delta}}^{(0)}, \bm{y}^{(0)})}{\partial y_t^{(0)}}  \\
    & = - \frac{\partial \phi(\bm{x}_t^{(0)},\widehat{\bm{\delta}}^{(0)})}{\partial(\widehat{\bm{\delta}}^{(0)})^{\top}} \left\{\frac{\partial \bm{g}(\widehat{\bm{\delta}}^{(0)}, \bm{y}^{(0)})}{\partial (\widehat{\bm{\delta}}^{(0)})^{\top}}\right\}^{-1} \frac{\partial \phi(\bm{x}_t^{(0)},\widehat{\bm{\delta}}^{(0)})}{\partial \widehat{\bm{\delta}}^{(0)}},
\end{align*}
where 
\begin{align*}
   \frac{\partial \bm{g}(\widehat{\bm{\delta}}^{(0)}, \bm{y}^{(0)})}{\partial (\widehat{\bm{\delta}}^{(0)})^{\top}}  =  \sum_{i=1}^{n_0} \Big[ -\frac{\partial \phi(\bm{x}_i^{(0)},\widehat{\bm{\delta}}^{(0)})}{\partial \widehat{\bm{\delta}}^{(0)}}  \frac{\partial \phi(\bm{x}_i^{(0)},\widehat{\bm{\delta}}^{(0)})}{\partial(\widehat{\bm{\delta}}^{(0)})^{\top}} + \{y_i^{(0)} - \phi(\bm{x}_i^{(0)}, \widehat{\bm{\delta}}^{(0)})\} \frac{\partial^2 \phi(\bm{x}_i^{(0)}, \widehat{\bm{\delta}}^{(0)})}{\partial \widehat{\bm{\delta}}^{(0)} \partial (\widehat{\bm{\delta}}^{(0)})^{\top}} \Big]. 
\end{align*}

\subsection{Spatial autoregressive model}\label{calcu:sar}

In this section, we derive the explicit expressions of $\partial \widehat{\bm{\mu}}^{(0)}/\partial (\bm{y}^{(0)})^{\top}$ for two commonly used estimation approaches, which are required to compute the GTLMMA criterion. In SAR models, parameter estimation may be obtained via maximum likelihood estimation (MLE), two-stage least squares (2SLS), or quasi-maximum likelihood estimation (QMLE) \citep{ord1975estimation, kelejian1998, lee2004asymptotic}. We derive the expressions first for MLE and then for 2SLS; the corresponding results for QMLE follow analogously. For the SAR model, the covariance matrix of the working mean is given by $\bm{\Omega}_0=\sigma_0^2 (\bm{I}_{n_0} - \rho^{(0)} \bm{W}_0)^{-1}\{(\bm{I}_{n_0} - \rho^{(0)} \bm{W}_0)^{-1}\}^{\top}$, and $\widehat{\bm{\Omega}}_0$ denotes the corresponding plug-in estimator obtained by replacing $\sigma_0$ and $\rho^{(0)}$ with their estimates.
\begin{enumerate}
    \item \textbf{MLE}: We assume that the parameter estimators $\widehat{\rho}^{(k)}$, $\widehat{\bm{\beta}}^{(k)}$, and $\widehat{\sigma}_k^2$ are all Maximum Likelihood estimates. For $k=0,\dots,K$, define $\bm{S}_k=\bm{I}_{n_k} - \rho^{(k)} \bm{W}_k$, and the log likelihood function of $\bm{y}^{(k)}$ is
\begin{align*}
     \log \mathcal{L}(\bm{y}^{(k)},\rho^{(k)},\bm{\beta}^{(k)},\sigma_k^2) 
    =  -\frac{n_k\log(2\pi \sigma_k^2) }{2} + \frac{1}{2} \log |\bm{S}_k|^2 
    - \frac{\| \bm{S}_k \bm{y}^{(k)}  - \bm{X}^{(k)}\bm{\beta}^{(k)}  \|^2}{2\sigma_k^2}.
\end{align*}
Define $\widehat{\bm{S}}_k=\bm{I}_{n_k} - \widehat{\rho}^{(k)} \bm{W}_k$ and $\bm{A}_k=\bm{I}_{n_k}-\bm{X}^{(k)}((\bm{X}^{(k)})^{\top}\bm{X}^{(k)})^{-1}(\bm{X}^{(k)})^{\top}$, then $\widehat{\bm{\beta}}^{(k)}=((\bm{X}^{(k)})^{\top}\bm{X}^{(k)})^{-1}(\bm{X}^{(k)})^{\top}\widehat{\bm{S}}_k\bm{y}^{(k)}$ and $\widehat{\sigma}_k^2=\|\widehat{\bm{S}}_k \bm{y}^{(k)}  - \bm{X}^{(k)} \widehat{\bm{\beta}}^{(k)}\|^2/n_k=\|\bm{A}_k \widehat{\bm{S}}_k \bm{y}^{(k)}\|^2/n_k$. It is seen that
\begin{align*}
    & \log \mathcal{L}(\bm{y}^{(k)},\rho^{(k)},\widehat{\bm{\beta}}^{(k)},\widehat{\sigma}_k^2) \\
    = & -\frac{n_k\log(2\pi \widehat{\sigma}_k^2) }{2} + \frac{1}{2}  \log | \widehat{\bm{S}}_k|^2 
    - \frac{\| \widehat{\bm{S}}_k \bm{y}^{(k)}  - \bm{X}^{(k)} \widehat{\bm{\beta}}^{(k)}  \|^2}{2\widehat{\sigma}_k^2} \\
    = & \frac{n_k \{\log(n_k)-\log(2\pi)-1\}}{2}-\frac{n_k}{2} \log \{ \| \bm{A}_k \widehat{\bm{S}}_k \bm{y}^{(k)} \|^2 \} +  \frac{1}{2} \log |\widehat{\bm{S}}_k|^2.
\end{align*}
Setting
\[
\frac{\partial}{\partial \rho}
\log \mathcal{L}\bigl(\bm{y}^{(k)}, \rho, \widehat{\bm{\beta}}^{(k)}(\rho), \widehat{\sigma}_k^2(\rho)\bigr)
\Big|_{\rho=\widehat{\rho}^{(k)}} = 0
\]
yields
\begin{eqnarray*}
 0 =   n_k (\bm{y}^{(k)})^{\top} \bm{W}_k^{\top} \bm{A}_k \left( \bm{I}_{n_k}- \widehat{\rho}^{(k)}\bm{W}_k \right)  \bm{y}^{(k)} - \mathrm{tr} \left\{ (\widehat{\bm{S}}_k)^{-1} \bm{W}_k \right\} \| \bm{A}_k \widehat{\bm{S}}_k \bm{y}^{(k)} \|^2.
\end{eqnarray*}
Taking the derivative with respect to $\bm{y}^{(0)}$ for the above formula when $k=0$, we have
\begin{eqnarray*}
    \bm{0} &=& n_0 ( \bm{W}_0^{\top} \bm{A}_0 + \bm{A}_0\bm{W}_0  - 2  \widehat{\rho}^{(0)} \bm{W}_0^{\top} \bm{A}_0 \bm{W}_0  ) \bm{y}^{(0)} \\
    && - n_0  (\bm{y}^{(0)})^{\top} \bm{W}_0^{\top} \bm{A}_0 \bm{W}_0 \bm{y}^{(0)} \frac{\partial \widehat{\rho}^{(0)}}{\partial \bm{y}^{(0)}}  - \| \bm{A}_0 \widehat{\bm{S}}_0 \bm{y}^{(0)} \|^2 \mathrm{tr} ( \widehat{\bm{S}}_0^{-1}   \bm{W}_0\widehat{\bm{S}}_0^{-1} \bm{W}_0 ) \frac{\partial \widehat{\rho}^{(0)}}{\partial \bm{y}^{(0)}}     \\
    &&- 2 \mathrm{tr} ( \widehat{\bm{S}}_0^{-1} \bm{W}_0 ) \left\{ \widehat{\bm{S}}_0^{\top} \bm{A}_0  \widehat{\bm{S}}_0 \bm{y}^{(0)} +  (\bm{y}^{(0)})^{\top} \bm{W}_0^{\top} \bm{A}_0 \left( \widehat{\rho}^{(0)}\bm{W}_0 - \bm{I}_{n_0} \right) \bm{y}^{(0)} \frac{\partial \widehat{\rho}^{(0)}}{\partial \bm{y}^{(0)}} \right\}.
\end{eqnarray*}
Therefore, we can derive the following explicit form of $\partial \widehat{\rho}^{(0)}/\partial \bm{y}^{(0)}$ as 
\begin{eqnarray}\label{partial:rhoy}
    \frac{\partial \widehat{\rho}^{(0)}}{\partial \bm{y}^{(0)}} & = & \left[ n_0 \left\{ \bm{W}_0^{\top} \bm{A}_0 + \bm{A}_0\bm{W}_0  - 2  \widehat{\rho}^{(0)} \bm{W}_0^{\top} \bm{A}_0 \bm{W}_0  \right\}  - 2 \mathrm{tr} \left\{ \widehat{\bm{S}}_0^{-1} \bm{W}_0 \right\} \widehat{\bm{S}}_0^{\top} \bm{A}_0  \widehat{\bm{S}}_0 \right] \bm{y}^{(0)} \nonumber \\
    && \times \left[ n_0  (\bm{y}^{(0)})^{\top} \bm{W}_0^{\top} \bm{A}_0 \bm{W}_0 \bm{y}^{(0)}+ \| \bm{A}_0 \widehat{\bm{S}}_0 \bm{y}^{(0)} \|^2 \mathrm{tr} \left\{ \widehat{\bm{S}}_0^{-1}   \bm{W}_0\widehat{\bm{S}}_0^{-1} \bm{W}_0 \right\} \right. \nonumber \\
    && - \left. 2 \mathrm{tr} \left\{ \widehat{\bm{S}}_0^{-1} \bm{W}_0 \right\}  (\bm{y}^{(0)})^{\top} \bm{W}_0^{\top} \bm{A}_0 \widehat{\bm{S}}_0 \bm{y}^{(0)} \right]^{-1},
\end{eqnarray}
and we can further derive 
$$
\partial \widehat{\bm{\beta}}^{(0)}/\partial (\bm{y}^{(0)})^{\top}=((\bm{X}^{(0)})^{\top}\bm{X}^{(0)})^{-1}(\bm{X}^{(0)})^{\top}\left\{\widehat{\bm{S}}_0 -\bm{W}_0 \bm{y}^{(0)} \frac{\partial \widehat{\rho}^{(0)}}{\partial \bm{y}^{(0)}} \right\}.
$$
This formula also applies to the quasi-maximum likelihood estimation (QMLE) proposed by Lee (\citeyear{lee2004asymptotic}), which employs the same estimation procedure as MLE but remains valid even when the disturbance term is not normally distributed, differing from MLE only in its theoretical properties.

\item \textbf{2SLS}: Since the computation of the determinant $|\bm{S}_k|$ in the ML approach incurs a significant computational burden, especially with a large sample size, we instead employ the 2SLS approach in this section, which is computationally simpler and distribution-free. According to \citet{kelejian1998}, the instrumental variables matrix $\bm{H}^{(k)}$ in the 2SLS approach is composed of a subset of the linearly independent $p+1$ (or more) columns of $(\bm{X}^{(k)}, \bm{W}_k\bm{X}^{(k)}, \bm{W}_k^2\bm{X}^{(k)},\dots)$. In Section \ref{sim}, we present the specific form of $\bm{H}^{(k)}$ as $\bm{H}^{(k)} = (\bm{X}^{(k)}, \bm{W}_k \bm{X}^{(k)})$. Let $\bm{Z}_k=(\bm{X}^{(k)}, \bm{W}_k\bm{y}^{(k)})$, then the 2SLS estimator of $\bm{\delta}^{(k)}$ is
$\widehat{\bm{\delta}}_{2SLS}^{(k)}=\left[ \bm{Z}_k^{\top} \bm{Q}_k \bm{Z}_k  \right]^{-1} \bm{Z}_k^{\top} \bm{Q}_k \bm{y}^{(k)}$,
where $\bm{Q}_k=\bm{H}^{(k)} ((\bm{H}^{(k)})^{\top}\bm{H}^{(k)})^{-1}  (\bm{H}^{(k)})^{\top}$. For $k=0$ and $i=1,\dots,n_0$, we calculate that
\begin{eqnarray*}
\frac{\partial ( \bm{Z}_0^{\top} \bm{Q}_0 \bm{Z}_0 ) }{\partial y_i^{(0)}} & = & \frac{\partial \begin{pmatrix}
 (\bm{X}^{(0)})^{\top}\bm{Q}_0\bm{X}^{(0)}, & (\bm{X}^{(0)})^{\top}\bm{Q}_0\bm{W}_0\bm{y}^{(0)} \\
 (\bm{y}^{(0)})^{\top}\bm{W}_0^{\top}\bm{Q}_0\bm{X}^{(0)}, & (\bm{y}^{(0)})^{\top}\bm{W}_0^{\top}\bm{Q}_0\bm{W}_0\bm{y}^{(0)}
\end{pmatrix} }{\partial y_i^{(0)}}\\
& = & \begin{pmatrix}
\bm{0}_{p \times p}, & \{[\bm{W}_0^{\top}\bm{Q}_0\bm{X}^{(0)}]_i\}^{\top} \\   
[\bm{W}_0^{\top}\bm{Q}_0\bm{X}^{(0)}]_i, & 2[\bm{W}_0^{\top}\bm{Q}_0\bm{W}_0\bm{y}^{(0)}]_i 
\end{pmatrix},
\end{eqnarray*}
where $[\cdot]_i$ denotes the $i$th row of the corresponding matrix or the $i$th element of the corresponding vector. For simplicity, we do not list this notation in other places when it is used. Thus, 
\begin{eqnarray*}
    \frac{\partial \widehat{\bm{\delta}}_{2SLS}^{(0)}}{\partial (\bm{y}^{(0)})^{\top}}
    & = & - \left[ \bm{Z}_0^{\top} \bm{Q}_0 \bm{Z}_0  \right]^{-1} \frac{\partial ( \bm{Z}_0^{\top} \bm{Q}_0 \bm{Z}_0 ) }{\partial (\bm{y}^{(0)})^{\top}}  \left[ \bm{Z}_0^{\top} \bm{Q}_0 \bm{Z}_0  \right]^{-1} \\
    && \times \bm{Z}_0^{\top} \bm{Q}_0 \bm{y}^{(0)} + \left[ \bm{Z}_0^{\top} \bm{Q}_0 \bm{Z}_0  \right]^{-1} \frac{\partial ( \bm{Z}_0^{\top} \bm{Q}_0 \bm{y}^{(0)} ) }{\partial (\bm{y}^{(0)})^{\top}},
\end{eqnarray*}
where 
\begin{eqnarray*}
\frac{\partial ( \bm{Z}_0^{\top} \bm{Q}_0 \bm{Z}_0 ) }{\partial (\bm{y}^{(0)})^{\top} } = \left[\frac{\partial ( \bm{Z}_0^{\top} \bm{Q}_0 \bm{Z}_0 ) }{\partial y_1^{(0)}},\dots, \frac{\partial ( \bm{Z}_0^{\top} \bm{Q}_0 \bm{Z}_0 ) }{\partial y_{n_0}^{(0)}}\right],
\end{eqnarray*}
and
\begin{eqnarray*}
\frac{\partial ( \bm{Z}_0^{\top} \bm{Q}_0 \bm{y}^{(0)} ) }{\partial (\bm{y}^{(0)})^{\top}}  = \frac{\partial \begin{pmatrix}
(\bm{X}^{(0)})^{\top}\bm{Q}_0\bm{y}^{(0)}  \\
 (\bm{y}^{(0)})^{\top}\bm{W}_0^{\top}\bm{Q}_0\bm{y}^{(0)} 
\end{pmatrix} }{\partial (\bm{y}^{(0)})^{\top}} =
\begin{pmatrix}
 (\bm{X}^{(0)})^{\top}\bm{Q}_0 \\
(\bm{y}^{(0)})^{\top} ( \bm{W}_0^{\top}\bm{Q}_0 +  \bm{Q}_0^{\top}\bm{W}_0)
\end{pmatrix}.
\end{eqnarray*}
Overall, $\partial \widehat{\rho}_{2SLS}^{(0)}/\partial \bm{y}^{(0)}$ is given by the last row of this matrix, and the derivative of $\partial \widehat{\bm{\beta}}_{2SLS}^{(0)}/\partial (\bm{y}^{(0)})^{\top}$ is given by its first $p$ rows.

\end{enumerate}

\subsection{Semiparametric model} \label{calcu:semi}

    In this section, we provide the formal model specification and spline-based construction underlying the semiparametric GTLMMA extension. 
    
    For $k=0,\dots,K$, the semiparametric model is specified as 
    \begin{align*}
        \bm{y}^{(k)} = \bm{Z}^{(k)}\bm{\delta}^{(k)} + \bm{g}^{(k)}(\bm{V}^{(k)}) + \bm{e}^{(k)}.
    \end{align*}
    The conditional mean function is given by $\bm{\mu}^{(k)}_{\mathrm{wm}} = \bm{Z}^{(k)}\bm{\delta}^{(k)} + \bm{g}^{(k)}(\bm{V}^{(k)})$, which consists of two components: A parametric component $\bm{Z}^{(k)}\bm{\delta}^{(k)}$ with an unknown parameter $\bm{\delta}^{(k)} \in \mathbb{R}^p$, and a nonparametric component $\bm{g}^{(k)}(\bm{V}^{(k)})=(g^{(k)}(\bm{v}_{1}^{(k)}),\dots,g^{(k)}(\bm{v}_{n_k}^{(k)}))^{\top}$. In the additive partially linear model, the nonparametric component adopts a commonly used additive structure, $g^{(k)}(\bm{v}_{i}^{(k)})=\sum_{l=1}^{q_k} g_l^{(k)}(v_{il}^{(k)})$, where each $g_l^{(k)}(\cdot)$ is a one-dimensional smooth function.
    For $k=0,\dots,K$, assume that $\mathbb{E}(e_i^{(k)} \mid \bm{z}_i^{(k)}, \bm{v}_i^{(k)})=0$ and $\mathbb{E}\{(e_i^{(k)})^2 \mid \bm{z}_i^{(k)}, \bm{v}_i^{(k)}\}=(\sigma_i^{(k)})^2$.

    When dealing with semiparametric models, we approximate the additive nonparametric component $\bm{g}^{(k)}(\bm{V}^{(k)})$ using B-spline basis expansions, following \citet{hu2023optimal}, thereby yielding a linear representation that enables ordinary least squares estimation. Specifically, for each model $k$, each nonparametric component $g_l^{(k)}(v_{il}^{(k)})$ is approximated by a linear combination of normalized B-spline basis functions, $B_l^{(k)}(v) = [b_{l1}(v), \ldots, b_{lt_l^{(k)}}(v)]^{\top}$, where $t_l^{(k)} = r_l^{(k)} + S_l^{(k)}$ denotes the total number of basis functions determined by the spline degree $r_l^{(k)}$ and the number of interior knots $S_l^{(k)}$. Accordingly, we approximate $g_l^{(k)}(v_{il}^{(k)})$ by
    $\{B_l^{(k)}(v_{il}^{(k)})\}^{\top} \bm{\gamma}_l^{(k)}$, where $\bm{\gamma}_l^{(k)}$ is the corresponding coefficient vector. The complete design matrix $\bm{F}^{(k)}$ for model $k$ is then constructed by horizontally concatenating the parametric covariates and the evaluated B-spline basis functions across all observations, yielding an $n_k \times (p + \sum_{l=1}^{q_k} t_l^{(k)})$ matrix:
    \begin{align*}
    \bm{F}^{(k)} = \begin{bmatrix}
    (\bm{z}_1^{(k)})^{\top} & \{B_1^{(k)}(v_{11}^{(k)})\}^{\top} & \cdots & \{B_{q_k}^{(k)}(v_{1q_k}^{(k)})\}^{\top} \\
    (\bm{z}_2^{(k)})^{\top} & \{B_1^{(k)}(v_{21}^{(k)})\}^{\top} & \cdots & \{B_{q_k}^{(k)}(v_{2q_k}^{(k)})\}^{\top} \\
    \vdots & \vdots & \ddots & \vdots \\
    (\bm{z}_{n_k}^{(k)})^{\top} & \{B_1^{(k)}(v_{n_k1}^{(k)})\}^{\top} & \cdots & \{B_{q_k}^{(k)}(v_{n_kq_k}^{(k)})\}^{\top}
    \end{bmatrix}.
    \end{align*}

    For each $k = 0, \dots, K$, the full parameter vector is $\bm{\theta}^{(k)} = [(\bm{\delta}^{(k)})^{\top}, (\bm{\gamma}_1^{(k)})^{\top}, \ldots, (\bm{\gamma}_{q_k}^{(k)})^{\top}]^{\top}$ with ordinary least-squares estimator $\widehat{\bm{\theta}}^{(k)} = \{(\bm{F}^{(k)})^{\top} \bm{F}^{(k)}\}^{-1} (\bm{F}^{(k)})^{\top} \bm{y}^{(k)}$. Let $\widehat{\bm{\delta}}^{(k)}$ denote the parametric component of $\widehat{\bm{\theta}}^{(k)}$, and let $\widehat{\bm{\gamma}}^{(0)}=\bigl[(\widehat{\bm{\gamma}}_1^{(0)})^{\top},\ldots,(\widehat{\bm{\gamma}}_{q_0}^{(0)})^{\top}\bigr]^{\top}$ denote the spline coefficient vector estimated from the target data. Since only the parametric component is transferred, the target-side parameter vector associated with the $k$th candidate predictor is $\widehat{\bm{\theta}}_k^{(0)}=[(\widehat{\bm{\delta}}^{(k)})^{\top},\widehat{\bm{\gamma}}^{(0)})^{\top}]^{\top}$, $k=0,\dots,K$. Accordingly, the corresponding predictor of $\bm{\mu}^{(0)}$ is $\widehat{\bm{\mu}}^{(k)}=\bm{F}^{(0)}\widehat{\bm{\theta}}_k^{(0)}=\bm{Z}^{(0)}\widehat{\bm{\delta}}^{(k)}+\bm{B}^{(0)}\widehat{\bm{\gamma}}^{(0)}$. This representation coincides with the transfer predictor used in the main text.

\subsection{Linear mixed-effects model} \label{calcu:panel}

In this section, we extend GTLMMA to the panel data setting under a linear mixed-effects model. For $k=0,\dots,K$, the $k$th dataset consists of $n_k$ clusters, where the $i$th cluster contains $T_i^{(k)}$ observations and $N_k = \sum_{i=1}^{n_k} T_i^{(k)}$. 
Let $\bm{y}_i^{(k)} \in \mathbb{R}^{T_i^{(k)}}$ be the response vector, $\bm{X}_i^{(k)} \in \mathbb{R}^{T_i^{(k)} \times p}$ and $\bm{Z}_i^{(k)} \in \mathbb{R}^{T_i^{(k)} \times l_k}$ the design matrices for fixed and random effects, respectively. 
The linear mixed-effects model is
\begin{align*}
\bm{y}_i^{(k)} = \bm{X}_i^{(k)} \bm{\delta}^{(k)} + \bm{Z}_i^{(k)} \bm{\vartheta}_i^{(k)} + \bm{\epsilon}_i^{(k)}, \quad i = 1, \dots, n_k,
\end{align*}
where $\bm{\delta}^{(k)}$ is the fixed effects vector, $\bm{\vartheta}_i^{(k)} \sim N(\bm{0}, \bm{\Lambda}_k)$ the cluster-specific random effects, and $\bm{\epsilon}_i^{(k)} \mid (\bm{X}_i^{(k)}, \bm{Z}_i^{(k)}, \bm{\vartheta}_i^{(k)}) \sim N(\bm{0}, \sigma_k^2 \bm{I}_{T_i^{(k)}})$, with $\{\bm{\vartheta}_i^{(k)}\}_{i=1}^{n_k}$ independent across clusters and independent of $\{\bm{\epsilon}_i^{(k)}\}_{i=1}^{n_k}$. Stacking the data, we define $\bm{y}^{(k)} = \{(\bm{y}_1^{(k)})^{\top},\ldots,(\bm{y}_{n_k}^{(k)})^{\top}\}^{\top}$, $\bm{X}_{\mathrm{FE}}^{(k)} = \{(\bm{X}_1^{(k)})^{\top},\ldots,(\bm{X}_{n_k}^{(k)})^{\top}\}^{\top}$, $\bm{Z}_{\mathrm{RE}}^{(k)} = \mathrm{diag}(\bm{Z}_1^{(k)},\ldots,\bm{Z}_{n_k}^{(k)})$, $\bm{\vartheta}^{(k)} = \{(\bm{\vartheta}_1^{(k)})^{\top},\ldots,(\bm{\vartheta}_{n_k}^{(k)})^{\top}\}^{\top}$ and $\bm{\epsilon}^{(k)}=\{(\bm{\epsilon}_1^{(k)})^\top,\ldots,
(\bm{\epsilon}_{n_k}^{(k)})^\top\}^\top$. Then the model can be written compactly as
\begin{align*}
\bm{y}^{(k)} = \bm{\mu}^{(k)} + \bm{\epsilon}^{(k)} = \bm{X}_{\mathrm{FE}}^{(k)} \bm{\delta}^{(k)} + \bm{Z}_{\mathrm{RE}}^{(k)} \bm{\vartheta}^{(k)} + \bm{\epsilon}^{(k)},
\end{align*}
where $\bm{\vartheta}^{(k)} \sim N(\bm{0},\bm{G}_k)$ and $\bm{G}_k = \mathrm{diag}(\bm{\Lambda}_k,\ldots,\bm{\Lambda}_k)$. Define $\bm{\Sigma}_k=\sigma_k^2\bm{I}_{N_k}+\bm{Z}_{\mathrm{RE}}^{(k)}\bm{G}_k(\bm{Z}_{\mathrm{RE}}^{(k)})^{\top}$.

For $k = 0, \dots, K$, define $\widehat{\bm{G}}_k$ as an estimator of $\bm{G}_k$ and $\widehat{\bm{\Sigma}}_k$ as an estimator of $\bm{\Sigma}_k$ given by $\widehat{\bm{\Sigma}}_k= \sigma_k^2 \bm{I}_{N_k}+\bm{Z}_{\mathrm{RE}}^{(k)}\widehat{\bm{G}}_k (\bm{Z}_{\mathrm{RE}}^{(k)})^{\top}$. We consider transferring the estimated fixed effects from the source models. This is motivated by the fact that fixed effects represent structural covariate-response relationships that tend to be similar across populations, whereas random effects capture population-specific heterogeneity and are thus less appropriate for transfer. Therefore, under the $k$th linear mixed-effects model, $\bm{\mu}^{(0)}$ can be estimated by 
\begin{align*}
    \widehat{\bm{\mu}}^{(k)}_{\mathrm{panel}} = \bm{X}_{\mathrm{FE}}^{(0)} \widehat{\bm{\delta}}^{(k)} + \bm{Z}_{\mathrm{RE}}^{(0)} \widehat{\bm{\vartheta}}^{(0)},
\end{align*}
where $\widehat{\bm{\delta}}^{(k)} = \{(\bm{X}_{\mathrm{FE}}^{(k)})^{\top} \widehat{\bm{\Sigma}}_k^{-1} \bm{X}_{\mathrm{FE}}^{(k)}\}^{-1}(\bm{X}_{\mathrm{FE}}^{(k)})^{\top} \widehat{\bm{\Sigma}}_k^{-1} \bm{y}^{(k)}$ and  $\widehat{\bm{\vartheta}}^{(0)} = \widehat{\bm{G}}_0 (\bm{Z}_{\mathrm{RE}}^{(0)})^{\top} \widehat{\bm{\Sigma}}_0^{-1}(\bm{y}^{(0)} - \bm{X}_{\mathrm{FE}}^{(0)} \widehat{\bm{\delta}}^{(0)})$. Here $\widehat{\bm{\vartheta}}^{(0)}$ is shared across all source models. The model averaging prediction is $\widehat{\bm{\mu}}_{\mathrm{panel}}(\bm{\omega}) = \sum_{k=0}^K \omega_k\widehat{\bm{\mu}}^{(k)}_{\mathrm{panel}}$. Let $\widehat{\bm{V}}_0 
= \widehat{\bm{\Sigma}}_0^{-1/2} \bm{X}_{\mathrm{FE}}^{(0)}
\{(\bm{X}_{\mathrm{FE}}^{(0)})^{\top} \widehat{\bm{\Sigma}}_0^{-1} \bm{X}_{\mathrm{FE}}^{(0)}\}^{-1} (\bm{X}_{\mathrm{FE}}^{(0)})^{\top} \widehat{\bm{\Sigma}}_0^{-1/2}$, which is symmetric and idempotent. First, we consider the case where the within-cluster variance $\sigma_0^2$ is known. Define $\widehat{\bm{P}}_0= \bm{I}_{N_0}-\sigma_0^2\widehat{\bm{\Sigma}}_0^{-1/2}(\bm{I}_{N_0} - \widehat{\bm{V}}_0 )\widehat{\bm{\Sigma}}_0^{-1/2}$ and $\widehat{\bm{A}}_0=\bm{I}_{N_0}- \widehat{\bm{P}}_0$. Then the fitted mean can be written as $\widehat{\bm{\mu}}^{(0)}_{\mathrm{panel}} = \widehat{\bm{P}}_0 \bm{y}^{(0)}$, which is a linear function of $\bm{y}^{(0)}$.

Let $\bm{\upsilon}^{(0)} = (\upsilon_1^{(0)}, \ldots, \upsilon_{J_0}^{(0)})^{\top}$ denote the $J_0 \times 1$ vector containing all unknown variance parameters in $\bm{\Lambda}_0$, and let $\widehat{\bm{\upsilon}}^{(0)}$ be its maximum likelihood estimator. We treat $\bm{\Sigma}_0$ as a function of the variance parameters 
$\bm{\upsilon}^{(0)}$. The panel GTLMMA criterion is
\begin{align}\label{panel_cri1}
\mathcal{C}_{\mathrm{panel}}(\bm{\omega})=\|  \bm{y}^{(0)} - \widehat{\bm{\mu}}_{\mathrm{panel}}(\bm{\omega}) \|^2 + 2 \omega_0 \sigma_0^2  \left\{ \mathrm{tr}(\widehat{\bm{P}}_0) + \sum_{j=1}^{J_0} \frac{\partial \widehat{\upsilon}_{j}^{(0)}}{\partial (\bm{y}^{(0)})^{\top}} \widehat{\bm{A}}_0 \left. \frac{\partial \bm{\Sigma}_0}{\partial \upsilon_j^{(0)}} \right|_{\upsilon_j^{(0)} = \widehat{\upsilon}_j^{(0)}} \widehat{\bm{A}}_0 \bm{y}^{(0)} \right\}.
\end{align}
Following \citet{Greven2010} and \citet{zhang2014biometrika}, the derivatives of the variance parameters in \eqref{panel_cri1} can be computed using a reduced-parameter representation that handles parameters on the boundary, and we omit the detailed algebra for brevity.

In practice, the minimization of $\mathcal{C}_{\mathrm{panel}}(\bm{\omega})$ is infeasible owing to unknown $\sigma_0$. Replacing $\sigma_0$ with its estimator $\widehat{\sigma}_0$ in the original estimators 
$\widehat{\bm{\Sigma}}_k$, $\widehat{\bm{\delta}}^{(k)}$, $\widehat{\bm{\vartheta}}^{(0)}$, 
$\widehat{\bm{\mu}}^{(k)}_{\mathrm{panel}}$, $\widehat{\bm{\mu}}_{\mathrm{panel}}(\bm{\omega})$, $\widehat{\bm{P}}_0$ and $\widehat{\bm{A}}_0$ yields the feasible counterparts 
$\check{\bm{\Sigma}}_k$, $\check{\bm{\delta}}^{(k)}$, $\check{\bm{\vartheta}}^{(0)}$, 
$\check{\bm{\mu}}_{\mathrm{panel}}^{(k)}$, $\check{\bm{\mu}}_{\mathrm{panel}}(\bm{\omega})$, $\check{\bm{P}}_0$ and $\check{\bm{A}}_0$ respectively.
Substituting these feasible estimators into $\mathcal{C}_{\mathrm{panel}}(\bm{\omega})$ then yields a feasible version of the criterion.
\begin{align*}
\widehat{\mathcal{C}}_{\mathrm{panel}}(\bm{\omega})=\|\bm{y}^{(0)}-\check{\bm{\mu}}_{\mathrm{panel}}(\bm{\omega})\|^2 + 2 \omega_0 \widehat{\sigma}_0^2  \left\{ \mathrm{tr}(\check{\bm{P}}_0) + \sum_{j=1}^{J_0} \frac{\partial \check{\upsilon}_{j}^{(0)}}{\partial (\bm{y}^{(0)})^{\top}} \check{\bm{A}}_0 \left. \frac{\partial \bm{\Sigma}_0}{\partial \upsilon_j^{(0)}} \right|_{\upsilon_j^{(0)} = \check{\upsilon}_j^{(0)}} \check{\bm{A}}_0 \bm{y}^{(0)} \right\}.
\end{align*}
The optimal weight vector is $\widehat{\bm{\omega}} = \arg\min_{\bm{\omega} \in \mathcal{W}} \widehat{\mathcal{C}}_{\mathrm{panel}}(\bm{\omega})$.

\subsection{Nonparametric regression methods} \label{calcu:nonparametric}

    This section provides additional details for the nonparametric extension discussed in Section~\ref{exten:non}. Many nonparametric methods have been proposed for estimating unknown regression functions, including neural networks \citep{bauer2019deep,shen2025Engression}, $K$-nearest neighbors \citep{demirkaya2024optimal,matabuena2024conditional}, random forests \citep{breiman2001random}, local polynomial regression \citep{fan1995local,welsh1996robust}, and related methods; see \citet{hastie2009elements} for a comprehensive overview. We focus on random forests as a representative example and then briefly discuss how the same idea can be extended to other nonparametric estimators.

    To construct the random forest predictor, for $k=0,\dots,K$, we fit a forest using $(\bm{X}^{(k)},\bm{y}^{(k)})$ and evaluate the fitted predictor at the target covariates $\bm{X}^{(0)}$. Each forest consists of $M_k$ regression trees built using the classification and regression trees (CART) algorithm. For $m=1,\dots,M_k$, the prediction from the $m$th tree for a target observation $\bm{x}_i^{(0)}$ admits the linear representation
    \begin{align*}
        \widehat{y}_i^{(m,k)} = \bm{p}_{\mathrm{BL}(m,k)}^{\top}(\bm{x}_i^{(0)}, \bm{X}^{(k)}, \bm{y}^{(k)}, \mathscr{B}_{(m,k)}, \bm{\Theta}_{(m,k)}) \bm{y}^{(k)},
    \end{align*}
    where $\bm{p}_{\mathrm{BL}(m,k)}(\cdot)$ is an $n_k$-dimensional weight vector determined by the tree structure. Here $\mathscr{B}_{(m,k)}$ denotes the bootstrap sample used to grow the $m$th tree, and $\bm{\Theta}_{(m,k)}$ collects the randomization parameters, such as feature subsampling and split selection. 
    
    Aggregating the tree-level predictions yields the forest predictor $\widehat{y}_i^{(k)}(\bm{\pi}^{(k)}) = \sum_{m=1}^{M_k} \pi_{m}^{(k)} \widehat{y}_i^{(m,k)}$, where $\bm{\pi}^{(k)}=(\pi_{1}^{(k)},\dots,\pi_{M_k}^{(k)})^\top$ denotes the vector of tree weights with $\bm{\pi}^{(k)} \in \mathcal{W}_{\mathrm{forest}}^{(k)}$, and $\mathcal{W}_{\mathrm{forest}}^{(k)}=\{\bm{\pi}^{(k)}: \pi_{m}^{(k)} \ge 0,\ \sum_{m=1}^{M_k} \pi_{m}^{(k)} =1 \}$. The weights may be uniform, $\pi_{m}^{(k)} = M_k^{-1}$, corresponding to the standard random forest, or determined by the adaptive scheme of \citet{chen2024optimal}. Stacking the predictions over all $i=1,\dots,n_0$ gives
    \begin{align*}
        \widehat{\bm{\mu}}^{(k)}=\{\widehat{y}_1^{(k)}(\bm{\pi}^{(k)}),\dots,\widehat{y}_{n_0}^{(k)}(\bm{\pi}^{(k)})\}^{\top}=\bm{P}_{k,\pi}^{(0)} \bm{y}^{(k)},
    \end{align*}
    where $\bm{P}_{k,\pi}^{(0)}\in\mathbb{R}^{n_0\times n_k}$ is the data-dependent weight matrix induced by the fitted forest. It admits the decomposition $\bm{P}_{k,\pi}^{(0)}=\sum_{m=1}^{M_k}\pi_{m}^{(k)}\bm{P}_{\mathrm{BL}(m,k)}$, where $\bm{P}_{\mathrm{BL}(m,k)}$ denotes the matrix whose $i$th row equals $\bm{p}_{\mathrm{BL}(m,k)}^{\top}(\bm{x}_i^{(0)}, \bm{X}^{(k)}, \bm{y}^{(k)}, \mathscr{B}_{(m,k)}, \bm{\Theta}_{(m,k)})$. Thus, $\bm{P}_{k,\pi}^{(0)}$ is determined by the fitted tree structures and the aggregation weights $\bm{\pi}^{(k)}$.
    This linear representation motivates the use of $\bm{P}_{0,\pi}^{(0)}$ as a surrogate for the derivative term in the GTLMMA criterion.

    Existing transfer learning methods for random forests include several approaches. In the single-source setting, \citet{li2025transfer} fit a residual forest on the target to correct a source-trained forest, while \citet{segev2016learn} transfer knowledge by modifying tree structures and split thresholds. In the multi-source setting, \citet{xiang2024transfer} augment target features with predictions from source-trained forests and retrain a new forest. In contrast, GTLMMA combines source-specific and target-specific predictions through model averaging,  thereby providing a direct way to aggregate multiple sources.

    Other flexible nonparametric estimators, such as spline regression, kernel regression, or neural networks, can also be incorporated by transferring their fitted predictions $\widehat{\bm{\mu}}^{(k)}$ to the target sample. We omit implementation details for these alternatives for brevity. However, the core idea of GTLMMA can be extended to these methods under suitable conditions. For instance, in spline regression, one may draw on \citet{racine2023optimal}, where candidate estimators admit linear representations that enable Mallows-type weighting. In our setting, source-specific estimators are evaluated on the target covariates to produce $\widehat{\bm{\mu}}^{(k)}$, which are then combined within the GTLMMA framework. This suggests a possible route for extending the GTLMMA framework to nonparametric settings.

\section{Additional details on assumptions}   
\subsection{Additional assumptions for the semiparametric results}\label{sec:ass:semi}

This section introduces additional assumptions required for the semiparametric extension of the theoretical results.
Compared with the parametric specification, the semiparametric model involves an additional nonparametric component, which necessitates modifications to the existing assumptions. To accommodate this extension, we redefine the notation in several assumptions, as summarized in Table~\ref{semi:notation}.

\renewcommand{\arraystretch}{1.3}
\begin{table}[ht]
\centering
\small
  \caption{Comparison of the notation in assumptions between the parametric and semiparametric models}
 \footnotesize
{\begin{tabular}{|c|c|c|}
\hline
  \textbf{Assumptions} & \textbf{Parametric model} & \textbf{Semiparametric model} \\  \hline 
  Assumption \ref{assumption2} &  
\thead{$\bm{\mu}^{(0)}=f(\bm{X}^{(0)},\bm{\delta}^{(0)})$ if correctly \\ specified; otherwise $\mathbb{E}(\bm{y}^{(0)} \mid \bm{X}^{(0)})$}
  & \thead{$\bm{\mu}^{(0)} = \bm{Z}^{(0)}\bm{\delta}^{(0)} + \bm{g}^{(0)}(\bm{V}^{(0)})$ if correctly \\ specified;  otherwise $\mathbb{E}(\bm{y}^{(0)} \mid \bm{X}^{(0)} )$}
  \\  \hline
  \multirow{3}{*}{Assumption \ref{assumption4}} & $\widehat{\bm{\delta}}^{(k)}$ &  $\widehat{\bm{\theta}}_k^{(0)}$ (defined in \ref{calcu:semi}) \\ \cline{2-3} 
  ~ & $\bm{\delta}^{(k)*}$ & $\bm{\theta}_k^{(0)*}=[(\bm{\delta}^{(k)*})^{\top}, (\bm{\gamma}_1^{(0)*})^{\top}, \ldots, (\bm{\gamma}_{q_0}^{(0)*})^{\top}]^{\top}$ \\ \cline{2-3} 
  ~ & $\alpha_{k}$ & $\alpha_{k,\mathrm{semi}}$ \\ \hline
  \multirow{3}{*}{Assumption \ref{assumption5}} & $\bm{\delta}^{(k)*}$ & $\bm{\theta}_k^{(0)*}$ \\ \cline{2-3}
  ~ & $\bm{\Delta}^{(k)}$ & \thead{$\bm{\Theta}^{(k)}= \{\overline{\bm{\theta}}_k^{(0)}: \alpha_{k,\mathrm{semi}}^{-1/2} K^{-1/2}  \|\overline{\bm{\theta}}_k^{(0)}-\bm{\theta}_k^{(0)*} \|< c_3\},$\\where $c_3$ is a positive constant.} \\ \cline{2-3}
  ~ & $\overline{\bm{\mu}}^{(k)}=f(\bm{X}^{(0)},\overline{\bm{\delta}}^{(k)})$ with $\overline{\bm{\delta}}^{(k)} \in \bm{\Delta}^{(k)}$ & $\overline{\bm{\mu}}^{(k)}=\bm{F}^{(0)} \overline{\bm{\theta}}_k^{(0)}$ with $\overline{\bm{\theta}}_k^{(0)} \in \bm{\Theta}^{(k)}$ \\   \hline
  \multirow{2}{*}{Assumption \ref{assumption6}} & $\widehat{\bm{\mu}}^{(k)}=f(\bm{X}^{(0)},\widehat{\bm{\delta}}^{(k)})$ & $\widehat{\bm{\mu}}^{(k)}=\bm{F}^{(0)}\widehat{\bm{\theta}}_k^{(0)}$ (defined in \ref{calcu:semi}) \\ \cline{2-3}
  ~ & $\widetilde{\bm{\mu}}^{(k)}=f(\bm{X}^{(0)},\bm{\delta}^{(k)*})$ & $\widetilde{\bm{\mu}}^{(k)}=\bm{F}^{(0)}\bm{\theta}_k^{(0)*}$ \\ \hline
  \multirow{2}{*}{Assumption \ref{assumption7}} & $\widetilde{\bm{\mu}}_{\mathrm{new}}^{(k)}=f(\bm{X}_{\mathrm{new}}^{(0)},\bm{\delta}^{(k)*})$ & $\widetilde{\bm{\mu}}_{\mathrm{new}}^{(k)}=\bm{F}_{\mathrm{new}}^{(0)} \bm{\theta}_k^{(0)*}$ \\ \cline{2-3}
  ~ & $\overline{\bm{\mu}}_{\mathrm{new}}^{(k)}=f(\bm{X}_{\mathrm{new}}^{(0)},\overline{\bm{\delta}}^{(k)})$ with $\overline{\bm{\delta}}^{(k)} \in \bm{\Delta}^{(k)}$ & $\overline{\bm{\mu}}_{\mathrm{new}}^{(k)}=\bm{F}_{\mathrm{new}}^{(0)} \overline{\bm{\theta}}_k^{(0)}$ with $\overline{\bm{\theta}}_k^{(0)} \in \bm{\Theta}^{(k)}$ \\ \hline
\end{tabular}}
 \label{semi:notation}
\end{table}

The assumptions in Sections~\ref{theo:opti} and \ref{theo:weight} are interpreted under the semiparametric notation in Table~\ref{semi:notation}. In particular, the rate parameter $\alpha_k$ is replaced by $\alpha_{k,\mathrm{semi}}$ and $\bar{\alpha}_{\mathrm{semi}}$ in Assumptions~\ref{assumption4}, \ref{assumption5}, \ref{assumption6}, and \ref{assumption7}.

\begin{assumption}\label{ass:semi:eig}
Assume that $\lambda_{\max}\{(\bm Z^{(0)})^\top \bm Z^{(0)}/n_0\}=O_p(1)$, $\lambda_{\max}\{(\bm F^{(0)})^\top \bm F^{(0)}/n_0\}=O_p(1)$, and $\lambda_{\min}\{(\bm F^{(0)})^\top \bm F^{(0)}/n_0\}\ge c$ with probability tending to one for some constant $c>0$.
\end{assumption}
This condition ensures that the design matrices are well behaved and allows control of the derivative terms in the semiparametric setting.

\begin{assumption}\label{assumption10}
    \leavevmode
    \begin{enumerate}
        \item $\lambda_{\max}\{(\bm{Z}^{(0)})^{\top}\bm{Z}^{(0)}/n_0\}\le c_4$ almost surely for some constant $c_4>0$.
        \item Define $\overline{t}^{(0)}=\max_{1\le l\le q_0}t_l^{(0)}$. Assume that $(\overline{t}^{(0)})^{-2\varrho} = O( \{(  K\bar{\alpha}_{\mathrm{semi}}^{1/2}) \vee h \})$, where $\varrho > 0.5$ is a parameter measuring the smoothness of the function.
    \end{enumerate}    
\end{assumption}

Assumption~\ref{assumption10}(i) imposes an upper bound on the maximum eigenvalue of $(\bm{Z}^{(0)})^{\top}\bm{Z}^{(0)}/n_0$, which is a mild and common condition analogous to Condition (A1) of \citet{b27} and Condition (C.3) of \citet{Zhang2020JASA}.
Assumption~\ref{assumption10}(ii) restricts the growth of the spline basis dimension. By Lemma 8 of \citet{Stone1986AoS}, the spline approximation error is $O\{(\overline{t}^{(0)})^{-2\varrho}\}$, which is asymptotically dominated by $(K\bar{\alpha}_{\mathrm{semi}}^{1/2}) \vee h$.

\subsection{Discussion of assumptions for different model specifications}\label{dis:ass}

In this section, we provide a discussion of the assumptions underlying our main results, focusing on their plausibility and applicability across different model specifications. Specifically, we consider the parametric and semiparametric settings, and explain why the assumptions are reasonable in each case.

\begin{example}{(Linear regression model)} \label{ass:linear}
\begin{itemize}

    \item \textbf{Discussion of Assumption \ref{assumption3}:}  Since $\widehat{\bm{\mu}}^{(0)}= \bm{X}^{(0)} \widehat{\bm{\delta}}^{(0)}$, the derivative $\partial \widehat{\bm{\mu}}^{(0)}/\partial (\bm{y}^{(0)})^{\top}$ is given by the hat matrix $\bm{X}^{(0)} \{ (\bm{X}^{(0)})^{\top} \bm{X}^{(0)}\}^{-1} (\bm{X}^{(0)})^{\top}$ under ordinary least squares, and by $\bm{X}^{(0)} \{(\bm{X}^{(0)})^{\top} \bm{X}^{(0)} + \lambda \bm{I}_p\}^{-1} (\bm{X}^{(0)})^{\top}$ under ridge regression with penalty parameter $\lambda$. In both cases, the rank of the Jacobian is at most $p$, and its largest singular value is bounded by a constant. Therefore, Assumption \ref{assumption3} holds in these standard linear settings.

    \item \textbf{Discussion of Assumption \ref{assumption4}:} In linear regression models, when the parameter dimension $p$ diverges with the sample size $n_k$, the convergence rate of M-estimators is of order $p^{1/2}n_k^{-1/2}$ (\citet{Huber1973} requires $p^{2}/n_k \to 0$, while \citet{Portnoy1984} imposes $p(\log p)/n_k \to 0$).

    \item \textbf{Discussion of Assumption \ref{assumption5}:} Since $\overline{\bm{\mu}}^{(k)}=\bm{X}^{(0)} \overline{\bm{\delta}}^{(k)}$, we derive $\bm{U}_k=\bm{X}^{(0)}$ and 
    \begin{align*}
         \lambda_{\max}(\bm{U}_k^{\top}\bm{U}_k) 
     = \lambda_{\max}\{(\bm{X}^{(0)})^{\top}\bm{X}^{(0)}\}
     \le  \lambda_{\max} \left\{ \frac{(\bm{X}^{(0)})^{\top}\bm{X}^{(0)}}{n_0} \right\}n_0. 
    \end{align*}
    By imposing the condition $\lambda_{\max}\{ (\bm{X}^{(0)})^{\top}\bm{X}^{(0)}/n_0 \}=O_p(1)$, which is analogous to Condition (C.3) of \citet{Zhang2020JASA}, we obtain $\lambda_{\max}(\bm{U}_k^{\top}\bm{U}_k)=O_p(n_0)$. Therefore, Assumption \ref{assumption5} can be regarded as reasonable in the linear regression model.
    
\end{itemize}
\end{example}

\begin{example}{(Nonlinear regression model)} \label{ass:nonlinear}
\begin{itemize}

    \item \textbf{Discussion of Assumption \ref{assumption3}:}  The validity of Assumption \ref{assumption3} relies on some standard conditions commonly imposed in nonlinear least-squares estimation (see, e.g., \citet{Jennrich1969,White01061981,Feng03042022}). 
    In nonlinear regression settings, \citet{Feng03042022} control a related Jacobian complexity measure through a trace-type quantity. Assumption \ref{assumption3} instead directly bounds the rank and the largest singular value of the Jacobian, providing a sufficient high-level condition for controlling its magnitude. In particular, in settings where the Jacobian matrix $\partial \widehat{\bm{\mu}}^{(0)} / \partial (\bm{y}^{(0)})^\top$ is symmetric positive semidefinite, Assumption \ref{assumption3} implies $\mathrm{tr}\{\partial \widehat{\bm{\mu}}^{(0)}/\partial (\bm{y}^{(0)})^\top\} \le \mathrm{rank}(\bm{A}_0)\sigma_{\max}(\bm{A}_0)=O_p(p^{1/2}n_0^{1/2})$. \citet{Feng03042022} show under their Assumption 6(4) that a related trace-type quantity is of order $O_p(n_0^{1/2})$ in the fixed-$p$ setting. Under the same setting, Assumption \ref{assumption3} yields a trace bound of the same order.

    \item \textbf{Discussion of Assumption \ref{assumption4}:} In nonlinear regression models, when the parameter dimension is finite, the convergence rate of the estimators may be of order $n_k^{-1/2}$, as discussed in \citet{Jennrich1969} and \citet{white1982maximum}. When the parameter dimension $p$ diverges with $n_k$, typical convergence rates for nonlinear least squares estimators are slower. In particular, \citet{POLLARD2006548} and \citet{Feng03042022} indicate that the conventional rate assumption $p^{1/2} n_k^{-1/2}$ may still be too strong. Consequently, we introduce $\alpha_k$ to characterize the rate at which the parameter estimates converge to their pseudo-true values.

    \item \textbf{Discussion of Assumption \ref{assumption5}:} In nonlinear regression problems, it is often necessary to impose assumptions on the first-order derivatives of the regression function with respect to the parameters. In this paper, the derivative of the target model ($k=0$) is also related to the order of the term $\mathrm{tr}\{ \partial  \widehat{\bm{\mu}}^{(0)} /\partial (\bm{y}^{(0)})^{\top} \}$ in the criterion. Relevant constraints can be found in Assumption 6(3) of \citet{Feng03042022}.

\end{itemize}
\end{example}

\begin{example}{(SAR model)} \label{ass:SAR}
\begin{itemize}
    \item \textbf{Discussion of Assumption \ref{assumption3}:}
    By combining the explicit form of $\partial  \widehat{\bm{\mu}}^{(0)} /\partial (\bm{y}^{(0)})^{\top}$, Assumption \ref{assumption3} may require imposing certain restrictions on the eigenvalues of $\bm{W}_0$, $\widehat{\bm{S}}_0^{-1}$ and the order of $\bm{y}^{(0)}$. It is a reasonable condition. For example, we can assess its feasibility under MLE. We define 
    $$
    \widehat{\bm{P}}_k=(\bm{I}_{n_0} -  \widehat{\rho}^{(k)} \bm{W}_0)^{-1} \bm{X}^{(0)}\{(\bm{X}^{(k)})^{\top} \bm{X}^{(k)} \}^{-1}(\bm{X}^{(k)})^{\top}  (\bm{I}_{n_k} -  \widehat{\rho}^{(k)} \bm{W}_k)
    $$
    for $k \in \mathcal{K}$, and thus $\widehat{\bm{\mu}}^{(k)}= (\bm{I}_{n_0} -  \widehat{\rho}^{(k)} \bm{W}_0)^{-1}\bm{X}^{(0)}\widehat{\bm{\beta}}^{(k)}=\widehat{\bm{P}}_k\bm{y}^{(k)}$. By the chain rule, the Jacobian admits the decomposition 
    $$
    \partial \widehat{\bm{\mu}}^{(0)}/\partial (\bm{y}^{(0)})^\top = \widehat{\bm{P}}_0 + \{(\partial \widehat{\bm{P}}_0/\partial \widehat{\rho}^{(0)})\bm{y}^{(0)}\}\{\partial \widehat{\rho}^{(0)}/\partial \bm{y}^{(0)}\}^\top,
    $$ 
    where the second term is a rank-one matrix. Hence, $\mathrm{rank}\{\partial \widehat{\bm{\mu}}^{(0)}/\partial (\bm{y}^{(0)})^\top\} \le \mathrm{rank}(\widehat{\bm{P}}_0)+1 \le p+1$, verifying the rank condition in Assumption \ref{assumption3}. For the largest singular value, we have 
    $$
    \sigma_{\max}\{\partial \widehat{\bm{\mu}}^{(0)}/\partial (\bm{y}^{(0)})^\top\} \le \sigma_{\max}(\widehat{\bm{P}}_0) + \|(\partial \widehat{\bm{P}}_0/\partial \widehat{\rho}^{(0)})\bm{y}^{(0)}\|_2 \, \|\partial \widehat{\rho}^{(0)}/\partial \bm{y}^{(0)}\|_2.
    $$ 
    Under the conditions $\|\widehat{\bm{S}}_0\|=O_p(1)$ and $\|\widehat{\bm{S}}_0^{-1}\|=O_p(1)$, we have $\sigma_{\max}(\widehat{\bm{P}}_0)=O_p(1)$. Moreover, if $\|\bm{W}_0\|=O(1)$, $\|\bm{y}^{(0)}\|_2^2=O_p(n_0)$, and $|D_0| \ge c n_0^2$ with probability tending to one, where $D_0$ denotes the scalar denominator in the expression of $\partial \widehat{\rho}^{(0)}/\partial \bm{y}^{(0)}$ (see \eqref{partial:rhoy}), then $\|(\partial \widehat{\bm{P}}_0/\partial \widehat{\rho}^{(0)})\bm{y}^{(0)}\|_2=O_p(n_0^{1/2})$ and $\|\partial \widehat{\rho}^{(0)}/\partial \bm{y}^{(0)}\|_2=O_p(n_0^{-1/2})$. It follows that $\sigma_{\max}\{\partial \widehat{\bm{\mu}}^{(0)}/\partial (\bm{y}^{(0)})^\top\}=O_p(1)$. Therefore, Assumption \ref{assumption3} holds.
    
    \item \textbf{Discussion of Assumption \ref{assumption4}:} 
    Under appropriate regularity conditions, various estimators such as 2SLS, MLE, and QMLE yield consistent estimates $\widehat{\bm{\delta}}^{(k)}$ of the true parameter $\bm{\delta}^{(k)}$. In the fixed-dimensional case, these estimators are $\sqrt{n_k}$-consistent (Kelejian and Prucha, \citeyear{kelejian1998}; Lee, \citeyear{lee2004asymptotic}). In contrast, when the parameter dimension $p$ diverges as the sample size $n_k$ increases, the convergence rate becomes of order $n_k^{-1/2}p^{1/2}$ \citep{gupta2015inference,gupta2018pseudo}.

    \item \textbf{Discussion of Assumption \ref{assumption5}:} Define $\overline{\bm{S}}_k=\bm{I}_{n_0} - \overline{\rho}^{(k)} \bm{W}_0$, we derive $\overline{\bm{\mu}}^{(k)}=\overline{\bm{S}}_k^{-1}\bm{X}^{(0)} \overline{\bm{\beta}}^{(k)}$,  $\bm{U}_k=[\partial \overline{\bm{\mu}}^{(k)}/ \partial (\overline{\bm{\beta}}^{(k)})^{\top},\partial \overline{\bm{\mu}}^{(k)}/ \partial \overline{\rho}^{(k)}]=[\overline{\bm{S}}_k^{-1}\bm{X}^{(0)}, \overline{\bm{S}}_k^{-1}\bm{W}_0\overline{\bm{S}}_k^{-1}\bm{X}^{(0)}\overline{\bm{\beta}}^{(k)}]$ and 
    \begin{align}\label{eq2}
         \lambda_{\max}(\bm{U}_k^{\top}\bm{U}_k) 
     & \le \lambda_{\max}\{(\bm{X}^{(0)})^{\top}(\overline{\bm{S}}_k^{-1})^{\top}\overline{\bm{S}}_k^{-1}\bm{X}^{(0)}\} + \| (\bm{X}^{(0)})^{\top}(\overline{\bm{S}}_k^{-1})^{\top}\overline{\bm{S}}_k^{-1}\bm{W}_0\overline{\bm{S}}_k^{-1}\bm{X}^{(0)}\overline{\bm{\beta}}^{(k)} \| \nonumber \\
     & \quad +\lambda_{\max}\{(\overline{\bm{\beta}}^{(k)})^{\top}(\bm{X}^{(0)})^{\top}(\overline{\bm{S}}_k^{-1})^{\top} \bm{W}_0^{\top} (\overline{\bm{S}}_k^{-1})^{\top} \overline{\bm{S}}_k^{-1}  \bm{W}_0 \overline{\bm{S}}_k^{-1} \bm{X}^{(0)} \overline{\bm{\beta}}^{(k)}\} \nonumber \\
     & \le  \lambda_{\max}^2(\overline{\bm{S}}_k^{-1}) \lambda_{\max} \left\{ \frac{(\bm{X}^{(0)})^{\top}\bm{X}^{(0)}}{n_0} \right\}n_0 \nonumber \\
     & \quad +  \lambda_{\max}^3(\overline{\bm{S}}_k^{-1}) \lambda_{\max}(W_0) \lambda_{\max} \left\{ \frac{(\bm{X}^{(0)})^{\top}\bm{X}^{(0)}}{n_0} \right\}n_0 \|\overline{\bm{\beta}}^{(k)}\| \nonumber \\
     & \quad +  \lambda_{\max}^4(\overline{\bm{S}}_k^{-1}) \lambda_{\max}^2(W_0) \lambda_{\max} \left\{ \frac{(\bm{X}^{(0)})^{\top}\bm{X}^{(0)}}{n_0} \right\}n_0 \|\overline{\bm{\beta}}^{(k)}\|^2, 
    \end{align}
    where the first step is due to the inequality $\lambda_{\max}\{(\bm{A},\bm{B})^{\top}(\bm{A},\bm{B})\} \le \lambda_{\max}(\bm{A}^{\top}\bm{A}) + \lambda_{\max}(\bm{B}^{\top}\bm{B}) + \|\bm{A}^{\top}\bm{B}\|$, which holds for arbitrary matrices $\bm{A}$ and $\bm{B}$. It is seen that
    \begin{align}\label{eq3}
    \|\bm{\overline{\mu}}^{(k)}\|^2  & =   \| (\bm{I}_{n_0} - \overline{\rho}^{(k)} \bm{W}_0)^{-1} \bm{X}^{(0)}\bm{\overline{\beta}}^{(k)}\|^2 \nonumber \\
    & = (\bm{\overline{\beta}}^{(k)})^{\top}(\bm{X}^{(0)})^{\top} \{(\bm{I}_{n_0} - \overline{\rho}^{(k)} \bm{W}_0)^{-1}\}^{\top} (\bm{I}_{n_0} - \overline{\rho}^{(k)} \bm{W}_0)^{-1} \bm{X}^{(0)}\bm{\overline{\beta}}^{(k)} \nonumber \\
    & \ge \lambda_{\min}^2\{(\bm{I}_{n_0} - \overline{\rho}^{(k)} \bm{W}_0)^{-1}\} n_0\lambda_{\min}\{(\bm{X}^{(0)})^{\top}\bm{X}^{(0)}/n_0\} \| \bm{\overline{\beta}}^{(k)} \|^2.
    \end{align}

Analogous to Assumptions \ref{assumption2}(i) and (ii), we assume that 
$\bm{\overline{\mu}}^{(k)} = f(\bm{X}^{(0)}, \overline{\bm{\delta}}^{(k)})$ satisfies 
$\mathrm{E}\{n_0^{-1}\|\overline{\bm{\mu}}^{(k)}\|^2\} = O(1)$ for all $\overline{\bm{\delta}}^{(k)}$ in the neighborhood $\Delta^{(k)}$. In addition, we assume that $\lambda_{\min}\{(\bm{I}_{n_0} - \overline{\rho}^{(k)} \bm{W}_0)^{-1}\} >0$ 
and $\lambda_{\min}\{(\bm{X}^{(0)})^{\top}\bm{X}^{(0)}/n_0\} >0$ almost surely for $k=0,\dots,K$. Combining these conditions with \eqref{eq3}, we obtain $\|\bm{\overline{\beta}}^{(k)}\|^2 = O(1)$. Combining \eqref{eq2} with $\|\bm{\overline{\beta}}^{(k)}\|^2 = O(1)$ and some other standard conditions, such as $\lambda_{\max}(\overline{\bm{S}}_k^{-1})=O(1)$, $\lambda_{\max}(W_0)=O(1)$, and $\lambda_{\max}\{ (\bm{X}^{(0)})^{\top}\bm{X}^{(0)}/n_0 \}=O_p(1)$, we can further obtain $\lambda_{\max}(\bm{U}_k^{\top}\bm{U}_k)=O_p(n_0)$. These standard conditions are mild and analogous to Assumptions 3 and 4 in \citet{Z&Y2018}, and a more detailed discussion can be found in Section A.3 of \citet{Z&Y2018}. Taken together, Assumption \ref{assumption5} can be regarded as reasonable in the SAR context.
\end{itemize}
\end{example}

\begin{example}{(Semiparametric model)}

\begin{itemize}

    \item \textbf{Discussion of Assumption \ref{assumption4}:} In the semiparametric setting, Assumption~\ref{assumption4} is imposed on $\widehat{\bm{\theta}}_k^{(0)}$, which combines the source-side parametric estimator $\widehat{\bm{\delta}}^{(k)}$ and the target-side spline estimator $\widehat{\bm{\gamma}}^{(0)}$. The following discussion provides an intuitive justification of the convergence rate, based on standard semiparametric estimation results for $\widehat{\bm{\theta}}^{(k)}$. 
    The convergence rate of the parametric component has already been discussed in Example 2. According to \citet{Stone1985AoS}, the nonparametric component converges at the rate $n_k^{-\varrho/(2\varrho+1)}$ with $\varrho>0.5$, which is evidently slower than the parametric $n_k^{-1/2}$ rate. 
    Therefore, we relax the convergence rate assumption to $\{p + \sum_{l=1}^{q_k} (r_l^{(k)}+S_l^{(k)})\}^{1/2} n_k^{-1/2}$, where the term $p^{1/2} n_k^{-1/2}$ corresponds to the convergence rate of the parametric component, and $\{\sum_{l=1}^{q_k} (r_l^{(k)} + S_l^{(k)})\}^{1/2} n_k^{-1/2}$ corresponds to the convergence rate of the nonparametric component. According to \citet{Stone1985AoS}, $S_l^{(k)} \asymp n_k^{1/(2\varrho+1)}$, and hence $(S_l^{(k)})^{1/2} \cdot n_k^{-1/2} \asymp n_k^{1/(4\varrho+2)} \cdot n_k^{-1/2} \asymp n_k^{-\varrho/(2\varrho+1)}$, which coincides with the result in \citet{Stone1985AoS}. This supports the choice of the convergence rate in Assumption~\ref{assumption4}. A similar rate decomposition also applies to $\widehat{\bm{\theta}}_k^{(0)}$, where the parametric and nonparametric components arise from the source and target samples, respectively. For the target-side transferred vector $\widehat{\bm{\theta}}_k^{(0)}$, this leads to the rate $\alpha_{k,\mathrm{semi}}=p/n_k+(r_0-p)/n_0$, where $\widehat{\bm{\delta}}^{(k)}$ is estimated from the $k$th source sample and $\widehat{\bm{\gamma}}^{(0)}$ is estimated from the target sample.

    \item \textbf{Discussion of Assumption \ref{assumption5}:} Under the semiparametric notation in Table~\ref{semi:notation},  $\overline{\bm{\mu}}^{(k)}=\bm{F}^{(0)} \overline{\bm{\theta}}_k^{(0)}$. Hence, we derive $\bm{U}_k=\partial \overline{\bm{\mu}}^{(k)}/ \partial (\overline{\bm{\theta}}_k^{(0)})^{\top}=\bm{F}^{(0)}$ and 
    \begin{align*}
         \lambda_{\max}(\bm{U}_k^{\top}\bm{U}_k) 
     = \lambda_{\max}\{(\bm{F}^{(0)})^{\top}\bm{F}^{(0)}\}
     =  \lambda_{\max} \left\{ \frac{(\bm{F}^{(0)})^{\top}\bm{F}^{(0)}}{n_0} \right\}n_0. 
    \end{align*}
    By imposing condition $\lambda_{\max}\{ (\bm{F}^{(0)})^{\top}\bm{F}^{(0)}/n_0 \}=O_p(1)$, we have $\lambda_{\max}(\bm{U}_k^{\top}\bm{U}_k)=O_p(n_0)$. Therefore, Assumption \ref{assumption5} can be regarded as reasonable in the semiparametric model.
    
\end{itemize}
\end{example}

\bibliographystyle{econ-apa-like}  
\bibliography{references}   
\end{document}